\documentstyle{ar}
\begin{document}
\input{psfig}

\newcommand\ApJ{{\it Ap.\ J.\ }}
\newcommand\ApJL{{\it Ap.\ J.\ Lett.\ }}
\newcommand\ApJS{{\it Ap.\ J.\ Suppl.\ }}
\newcommand\PR{{\it Phys.\ Rev.\ }}
\newcommand\PRL{{\it Phys. Rev.\ Lett.\ }}
\newcommand\PL{{\it Phys.\ Lett.\ }}
\newcommand\MNRAS{{\it MNRAS\ }}
\newcommand\MNRASL{{\it MNRAS\ Lett.\ }}
\newcommand\AnA{{\it Astron.\ Astrophys.\ }}
\newcommand\BAAS{{\it Bull.\ Am.\ Astron.\ Soc.\ }}
\newcommand\NP{{\it Nucl.~Phys.\ }}
\newcommand\RMP{{\it Rev.\ Mod.\ Phys.\ }}
\newcommand\apj{\ApJ}
\newcommand\prl{\PRL}
\newcommand\ARAA{{\it ARAA}}

\newcommand\COBE{{\sl COBE}}

\def\VEV#1{\left\langle #1\right\rangle}
\def\sec{\ifmmode \,\, {\rm sec} \else sec \fi}
\def\eV {\ifmmode \,\, {\rm eV} \else eV \fi}
\def\keV{\ifmmode \,\, {\rm keV} \else keV \fi}
\def\MeV{\ifmmode \,\, {\rm MeV} \else MeV \fi}
\def\GeV{\ifmmode \,\, {\rm GeV} \else GeV \fi}
\def\TeV{\ifmmode \,\, {\rm TeV} \else TeV \fi}
\def\fm{\ifmmode \,\, {\rm fm} \else TeV \fi}
\def\pbarn{\ifmmode \,\, {\rm pb} \else pb \fi}
\def\km{\ifmmode {\rm km}\, \else km \fi}
\def\Mpc{\ifmmode {\rm Mpc}\, \else Mpc \fi}
\def\Gyr{\ifmmode {\rm Gyr}\, \else Gyr \fi}
\def\Mx{{m_{\chi}}}
\def\Mq{m_q}
\def\Msq{m_{\tilde q}}
\def\ra{\rightarrow}
\def\fun#1#2{\lower3.6pt\vbox{\baselineskip0pt\lineskip.9pt
  \ialign{$\mathsurround=0pt#1\hfil##\hfil$\crcr#2\crcr\sim\crcr}}}
\def\la{\mathrel{\mathpalette\fun <}}
\def\ga{\mathrel{\mathpalette\fun >}}
\def\order{{\cal O}}
\def\etal{{\rm et al.}}
\def\neut{{\tilde\chi}}
\def\mx{{m_{\chi}}}
\def\tanb{\tan\beta}
\def\Msf{ m_{\tilde f}}
\def\sbar#1{\kern 0.8pt
        \overline{\kern -0.8pt #1 \kern -0.8pt}
        \kern 0.8pt}  
\def\Nzsq{\VEV{Nz^2}}
\def\meter{\ifmmode \,\, {\rm m} \else m \fi}
\def\yr {\ifmmode \,\, {\rm yr} \else yr \fi}
\def\Ein{{E_{\rm in}}}
\def\sr{\ifmmode \,\, {\rm sr} \else sr \fi}
\def\sigann{(\sigma_A v)_{26}}
\def\kmsec{km sec$^{-1}$}
\def\Rf{\baselineskip=12pt\parindent=0pt \hangindent=3pc \hangafter=1}
\def\minim{{\rm min}}
\def\Msolar{M_\odot}
\def\hatn{{\bf \hat n}}
\def\hatm{{\bf \hat m}}
\def\hatk{{\bf \hat k}}
\def\veck{{\vec k}}
\def\vecx{{\vec x}}
\def\vecr{{\vec r}}

\def\slashchar#1{\setbox0=\hbox{$#1$}           
   \dimen0=\wd0                                 
   \setbox1=\hbox{/} \dimen1=\wd1               
   \ifdim\dimen0>\dimen1                        
      \rlap{\hbox to \dimen0{\hfil/\hfil}}      
      #1                                        
   \else					
      \rlap{\hbox to \dimen1{\hfil$#1$\hfil}}   
      /                                         
   \fi}

\long\def\comment#1{}

\title{The Cosmic Microwave Background and Particle
Physics\footnote{To appear in
\textit{Annu. Rev. Nucl. Part. Sci.} (1999)}}
\markboth{Marc Kamionkowski and Arthur Kosowsky}{The Cosmic
Microwave Background and Particle Physics}

\author{Marc Kamionkowski
\affiliation{Department of Physics, Columbia University, 538
West 120th Street, New York, New York 10027; e-mail: kamion@phys.columbia.edu}
Arthur Kosowsky
\affiliation{Department of Physics and Astronomy, Rutgers
University, Piscataway, New Jersey 08854-8019; e-mail: kosowsky@physics.rutgers.edu}}

\begin{keywords}
{cosmology, particle physics  \hfill CU-TP-935, RAP-246, astro-ph/9904108}

\end{keywords}

\begin{abstract}
In forthcoming years, connections between cosmology and
particle physics will be made increasingly important
with the advent of a new generation of cosmic microwave
background (CMB) experiments.  Here, we review a number of these 
links.  Our primary focus is on new CMB tests of inflation.  We
explain how the inflationary predictions for the geometry of the 
Universe and primordial density perturbations will be tested by
CMB temperature fluctuations, and how the gravitational waves
predicted by inflation can be pursued with the CMB polarization.
The CMB signatures of topological defects and primordial
magnetic fields from cosmological phase transitions are also
discussed.  
Furthermore, we review current and future CMB constraints on
various types of dark matter (e.g.\ massive neutrinos,
weakly interacting massive particles, axions, vacuum energy),
decaying particles, the baryon asymmetry of the Universe,
ultra-high-energy cosmic rays, exotic cosmological topologies,
and other new physics.
\end{abstract}

\maketitle

\section{Overview of the Cosmic Microwave Background}

In 1948, Alpher \& Hermann \cite{AlpHer49} realized that if
light elements were produced in a hot big bang, as Gamow and
others had suggested \cite{Gam46}, then the Universe today
should have a temperature of about 5 K.  When Penzias
\& Wilson discovered an anomalous background in 1964,
consistent with a blackbody spectrum at a temperature of $\sim3$
K \cite{PenWil65}, Dicke and his
collaborators immediately recognized it as the radiation
associated with this nonzero cosmological temperature
\cite{Dicetal65}.  Subsequent observations that confirm a
remarkable degree of isotropy (apart from a dipole
\cite{CorWil76,SmoGorMul77}, which can be interpreted as
our motion of $627\pm22$ km~s$^{-1}$ with respect to the
blackbody rest frame
\cite{Smoetal91,Smoetal92,Kogetal93,Fixetal94}) suggest an
extragalactic origin for this cosmic microwave background
(CMB).  Strong upper limits to any angular cross-correlation
between the CMB temperature and the extragalactic X-ray
background intensity \cite{Kneetal97,BouCriTur98} suggest that the CMB
comes from redshifts greater than those ($z\simeq2-4$) 
probed by
the active galactic nuclei and galaxy clusters that produce the
X-ray background.  This evidence, as well as the
exquisite blackbody spectrum
of the CMB \cite{Matetal94,Wrietal94,Fixetal96}, further supports
the notion that this radiation is the cosmological blackbody
postulated by Alpher \& Hermann.

Although they have a Planck spectrum, CMB photons are not in
thermal equilibrium.  The mean free path for scattering of
photons in the Universe must be huge, or else we would not see
galaxies and quasars out to distances of thousands of
Mpc.\footnote{Mpc$=3.3\times10^6$~light years$=3.09\times10^{24}$~cm.}
So where did these photons come from?  At early times ($t\la10^5$ y;
redshifts $z\ga1000$), the temperature of the Universe exceeded
an eV,
so the Universe consisted of a plasma of free electrons
and light nuclei.  CMB photons were tightly coupled to this plasma via 
Thomson scattering from the free electrons.  At a redshift of
$z\simeq1000$, the temperature dropped below a few eV, and
electrons and nuclei combined to form atoms.  At this point,
photons ceased interacting.  
A detailed analysis of ``recombination'' and 
the almost simultaneous (although slightly later) decoupling of
photons shows that CMB photons last scattered near a redshift of
$z\simeq1100$  \cite{JonWys76,KolTur90,Pee93}.

When we look at these CMB photons coming to us from all
directions in
the sky, we are therefore looking directly at a
spherical surface in the Universe that surrounds us at a
distance of $\sim10^4$ Mpc, as it was when the
Universe was only about 300,000 years old.  The temperature of
the CMB is found to be the same, to roughly one part in $10^5$, in 
every direction on the sky.  This remarkable
isotropy poses a fundamental
conundrum for the standard big-bang theory.  When these photons
last scattered, the size of a causally connected region of the
Universe was roughly 300,000 light years, and such a region
subtends an angle of only one degree on the sky.  Thus, when we
look at the CMB, we are looking at roughly 40,000 causally disconnected
regions of the Universe.  How is it, then, that each of these
has the same temperature to one part in $10^5$?  This is the
well-known isotropy, homogeneity, or horizon problem.

Another fundamental question in cosmology today is the origin
of the large-scale structure of the galaxy distribution.
The simplest and most plausible
explanation is that the observed inhomogeneities grew from tiny density
perturbations in the early Universe via gravitational instability.  
Mass from underdense
regions is drawn towards overdense regions, and in this way,
small primordial perturbations are amplified into the
structure we see in the
Universe today.  New support for this hypothesis was provided by 
the Cosmic Background Explorer (\COBE) detection of temperature
differences in the CMB of
roughly one part in $10^5$ \cite{Smoetal90}.
Heuristically, density perturbations induce
gravitational-potential perturbations at the surface of last
scatter;  photons that arrive from denser regions
climb out of deeper potential wells and thus appear redder
than those from underdense regions (the Sachs-Wolfe effect
\cite{SacWol67}).  Thus, the temperature fluctuations seen with
\COBE\ provide a snapshot of the tiny primordial perturbations
that gave rise to the large-scale structure we see in the
Universe today.  But this raises a second question:  If
large-scale structure grew via gravitational infall from tiny
inhomogeneities in the early Universe, where did these
primordial perturbations come from?

Before \COBE, there was no shortage of ideas for the origin of
large-scale structure, and, quite remarkably, all causal
mechanisms for producing
primordial perturbations have come from new ideas in particle theory:
primordial adiabatic perturbations from inflation
\cite{GutPi82,Haw82,Lin82b,Sta82,BarSteTur83},
late-time phase transitions \cite{Was86,HilSchFry89}, a
loitering Universe \cite{SahFelSte92}, scalar-field ordering
\cite{Vil82,Pre80}, topological defects \cite{ZelKozOku74,Kib76}
(such as cosmic strings \cite{Zel80,Vil81,SilVil84,Tur85},
domain walls \cite{HilSchFry89,PreRydSpe89}, textures
\cite{Tur89,Tur91}, or global monopoles \cite{BarVil89,BenRhi91}), 
superconducting cosmic strings \cite{Wit85,WitThoOst86},
isocurvature axion perturbations
\cite{SecTur85,TurWil91,AxeBraTur83,SteTur83,KofLin87,Lyt90,Lin91}, 
etc.

However, after \COBE, primordial
adiabatic perturbations (perturbations to the total density with 
equal fractional number-density perturbations in each species in the
Universe) seem to provide the only workable models.  Such
perturbations are produced naturally during inflation, a period
of exponential expansion in the early Universe driven by the
vacuum energy associated with some new scalar field
\cite{Gut81,Lin82a,AlbSte82}.
With adiabatic
perturbations, hotter regions at the surface of last scatter are
embedded in deeper potential wells, so the reddening due to
the gravitational redshift of the photons from these regions
partially cancels the higher intrinsic temperatures.  
When normalized to the amplitude of density perturbations indicated by galaxy
surveys, alternative models generically produce a larger
temperature fluctuation than that measured by \COBE\ 
\cite{KodSas86,EfsBon87,JafSteFri94}.  Recently, more detailed
calculations of the expected CMB-anisotropy amplitude have led
proponents of topological defects, the primary
alternative to inflation, to concede
that these models have difficulty accounting for the origin of
large-scale structure \cite{PenSelTur97,AlbBatRob97,Alletal97}.

Although inflation now seems to provide
the best candidate for the
origin of large-scale structure, 
the primary attraction of
inflation was originally that it
provided (and still does) the best (if not only) solution of
the horizon problem.  
For these reasons, inflation
has taken center stage in cosmology.  Although
inflation was for a long time speculative physics beyond the
realm of experimental tests, we are now entering a new era in
which the predictions of inflation will be tested with
unprecedented precision by CMB measurements.

The primary focus of this article is therefore to review
the predictions of inflation and how they will be tested with
the CMB.  Although inflation currently seems to provide the most 
promising paradigm for the origin of large-scale structure, it
is not yet well established.  Moreover, although the simplest
topological-defect models seem to be ruled out, it is still
certainly plausible that some more involved models may be
able to account for large-scale structure.  We therefore review
the CMB predictions of topological-defect models.  We also
discuss a number of other promising links between the CMB
and particle physics that do not necessarily have to do with the 
origin of structure, e.g.\ dark matter, neutrino properties,
decaying
particles, cosmological magnetic fields from early-Universe
phase transitions, parity violation, gravity theories, time
variation of fundamental parameters, and baryogenesis scenarios.

We are unfortunately unable to cover the larger bodies of
excellent work on the CMB in general, nor on the intersections
between particle physics and cosmology more generally.
Fortunately, a number of excellent reviews cover those subjects,
to which we cannot do justice here.
Lyth and Riotto \cite{LytRio99} review particle-physics models of
inflation;  Liddle and Lyth discuss structure formation
in inflation-inspired cold-dark-matter models.
Lidsey et al \cite{Lidetal97} review the production of density
perturbations and reconstruction of the inflaton potential from
the power spectra of density perturbations and gravitational
waves.  White, Scott, and Silk \cite{WhiScoSil94} review the CMB and structure
formation, and Hu and White \cite{HuWhi97b} provide a brief review of
the theory of CMB polarization.  Finally, see
References \cite{SunZel80,Rep95,Bir99} for reviews of the Sunyaev-Zeldovich
effect (the scattering of CMB photons from hot  gas in clusters
of galaxies), an intriguing and potentially very important
probe of the physics of clusters.

\section{Cosmic Microwave Background Observables}
\label{sec:cmbobservables}

\subsection{The Frequency Spectrum}
\label{sec:freqspectrum}

Standard cosmology predicts the CMB frequency spectrum to be
that of a perfect blackbody,
\begin{equation}
     S(\nu;T) = {2 h c^2 \nu^3 \over e^x-1},
\label{eq:blackbody}
\end{equation}
where $x=hc\nu/kT$, $h$ is Planck's constant, $c$ is the
velocity of light, $\nu$ is the frequency, $k$ is Boltzmann's
constant, and $T$ is the temperature.  Of the infinitude of
possible distortions to this spectrum, two common forms often
considered in the literature---Bose-Einstein and Compton
distortions---could arise from basic physical processes
before recombination.

If photons are released into the Universe from some nonthermal
process (e.g.\ decay of a massive particle) when the temperature 
of the Universe exceeds roughly 1 keV
(redshifts $z\ga10^6$ when 
the age of the Universe is $t\la10^7$ sec), they will come into
complete thermal equilibrium with the photons in the primordial
plasma.  More precisely, they attain kinetic equilibrium through
Compton scattering, double Compton scattering, and bremsstrahlung, 
and they attain 
chemical equilibrium (chemical potential $\mu=0$) because the
rate for photon-number-changing processes (e.g.\
$\gamma\gamma\ra\gamma\gamma\gamma$) that maintain a chemical
potential $\mu=0$ exceeds the expansion rate.  Therefore,
if any electromagnetic
energy is released into the Universe at such early times, it
will have no observable effect on the CMB.  However, if photons
are released at later times (but still before recombination),
they can distort the CMB frequency spectrum
\cite{SunZel80,DanDeZ77,Lig81,DanDeZ82,SarCoo83,FukKaw90,BurDanDeZ91a,BurDanDeZ91b,HuSil93b}.

\subsubsection{Bose-Einstein distortion}
Nonthermal photons produced in the redshift range $10^5
\la z  \la 3\times10^6$ (temperatures $T\simeq 0.1-1$ keV and
ages $t\simeq10^{7-9}$ sec) can still attain kinetic
equilibrium, but they will not attain chemical equilibrium, as
interactions that change the photon number occur less rapidly
than the expansion rate.  If electromagnetic energy
is released at these times, the CMB frequency dependence will be 
that of a Bose-Einstein gas with a nonzero chemical potential,
\begin{equation}
     S_\mu(\nu;T,\mu) = {2 h c^2 \nu^3 \over e^{x+\mu}-1},
\label{eq:boseeinstein}
\end{equation}
where $\mu$ is the (dimensionless) chemical potential.  The
Far Infrared Absolute Spectrophotometer (FIRAS) result for $\mu$
is $\mu=-1\pm10\times10^{-5}$ or a 95\%
confidence-level upper limit of $|\mu| < 9\times10^{-5}$
\cite{Fixetal96}.  It is possible that values of $\mu$ as small
as $10^{-6}$ could be probed by a future satellite mission
\cite{Shaetal95}.

\subsubsection{Compton distortion}
If photons are released at later times ($z\la10^5$) but still
before recombination ($z\simeq1100$; temperatures $T\simeq 1-100$
eV and times $t\simeq10^{9-13}$ sec), they do not have enough
time to come to either thermal or kinetic equilibrium and wind
up producing a ``Compton distortion'' of the form
\begin{equation}
     S_y(\nu;T,y)={2 h c^2 \nu^3 \over e^x-1}\left(1 + y x {1
     \over 1-e^{-x}} \left[x\coth(x/2)-4\right]\right),
\label{eq:comptony}
\end{equation}
to linear order in $y$
(the Kompaneets or ``Compton-$y$''
parameter).  If some CMB photons were rescattered after
recombination by a hot intergalactic gas, this would also
produce a Compton-$y$ distortion.  The FIRAS result for this
type of distortion is $y=-1\pm6\times10^{-6}$ or an upper limit
of $|y|<15\times10^{-6}$ at the 95\% confidence level
\cite{Fixetal96}.  The
consensus among the experimentalists we have surveyed seems to
be that it would be difficult to improve on this limit.

\subsection{Temperature and Polarization Power Spectra}

The primary aim of forthcoming CMB satellite
experiments, such as NASA's Microwave Anisotropy Probe (MAP) \cite{MAP} 
and the European Space Agency's Planck Surveyor \cite{Planck},
will be to map the temperature $T(\hatn)$ of the CMB and its
linear polarization, described by Stokes parameters $Q(\hatn)$
and $U(\hatn)$, as functions of position $\hatn$ on the sky.
Several temperature-polarization angular correlation 
functions, or equivalently, power spectra, can be 
extracted from such maps.
These quantities can be compared with detailed predictions from
cosmological models.

\subsubsection{Harmonic analysis for temperature anisotropies and
polarization}
{\it Temperature Anisotropies}\\  
The temperature map can be expanded in spherical harmonics,
\begin{equation}
     {T(\hatn) \over T_0} = 1 + \sum_{lm} a^{\rm T}_{(lm)}
     Y_{(lm)}(\hatn),
\label{Texpansion}
\end{equation}
where the mode amplitudes are given by
\begin{equation}
     a^{\rm T}_{(lm)}={1\over T_0}\int
     d\hatn\,T(\hatn)\,Y_{(lm)}^*(\hatn);
\label{Talms}
\end{equation}
this follows from orthonormality of the spherical harmonics.
Here, $T_0=2.728\pm0.002$ K is the cosmological mean CMB
temperature \cite{Fixetal96}.

{\it Linear polarization}\\  
The Stokes parameters (where $Q$ and $U$ are measured with
respect to the polar ${\bf \hat\theta}$ and azimuthal ${\bf \hat
\phi}$ axes) are components of a $2\times2$ symmetric
traceless tensor with two independent components,
\begin{equation}
  {\cal P}_{ab}(\hatn)={1 \over 2} \left( \begin{array}{cc}
   \vphantom{1\over 2}Q(\hatn) & -U(\hatn) \sin\theta \\
   -U(\hatn)\sin\theta & -Q(\hatn)\sin^2\theta \\
   \end{array} \right),
\label{eq:whatPis}
\end{equation}
where the subscripts $ab$ are tensor indices, and $Q$ and $U$
are given in temperature units.
Just as the temperature is expanded in terms of spherical
harmonics, the polarization tensor can be expanded
\cite{KamKosSte97a,KamKosSte97b,SelZal97,ZalSel97},
\begin{equation}
      {{\cal P}_{ab}(\hatn)\over T_0} =
      \sum_{lm} \left[ a_{(lm)}^{{\rm G}}Y_{(lm)ab}^{{\rm
      G}}(\hatn) +a_{(lm)}^{{\rm C}}Y_{(lm)ab}^{{\rm C}}(\hatn)
      \right],
\label{eq:Pexpansion}
\end{equation}
in terms of tensor spherical harmonics, $Y_{(lm)ab}^{\rm G}$
and $Y_{(lm)ab}^{\rm C}$.  It is well known that a vector field
can be decomposed into a curl (C) and a curl-free (gradient) (G)
part.
Similarly, a 
$2\times2$ symmetric traceless
tensor field can be decomposed into a tensor
analogue of a curl and a gradient part; the $Y_{(lm)ab}^{\rm G}$
and $Y_{(lm)ab}^{\rm C}$ form a complete orthonormal basis for the
``gradient'' (i.e.\ curl-free) and ``curl'' components of the
tensor field, respectively.\footnote{Our G and C are sometimes
referred to as the ``scalar'' and ``pseudo-scalar'' components
\cite{Ste96}, respectively, or with slightly different
normalization as E and B modes
\cite{ZalSel97} (although these should not be confused with the
radiation's electric- and magnetic-field vectors).}  Lengthy but
digestible expressions for the
$Y_{(lm)ab}^{\rm G}$ and $Y_{(lm)ab}^{\rm C}$ are given in terms 
of derivatives of spherical harmonics and also in terms of
Legendre functions in Reference \cite{KamKosSte97b}.
The mode amplitudes in Equation \ref{eq:Pexpansion} are given by
\begin{eqnarray}
     a^{\rm G}_{(lm)}&=&{1\over T_0}\int d\hatn\,{\cal P}_{ab}(\hatn)\, 
                                         Y_{(lm)}^{{\rm G}
					 \,ab\, *}(\hatn),\cr 
     a^{\rm C}_{(lm)}&=&{1\over T_0}\int d\hatn\,{\cal P}_{ab}(\hatn)\,
                                          Y_{(lm)}^{{\rm C} \,
					  ab\, *}(\hatn), 
\label{eq:Amplitudes}
\end{eqnarray}
which can be derived from the orthonormality properties of these 
tensor harmonics \cite{KamKosSte97b}.
Thus, given a polarization map ${\cal P}_{ab}(\hatn)$, the G and 
C components can be isolated by first carrying out the
transformations in Equation \ref{eq:Amplitudes} to the $a^{\rm
G}_{(lm)}$ and $a^{\rm C}_{(lm)}$ and then summing over the
first term on the right-hand side of Equation \ref{eq:Pexpansion}
to get the G component and over the second term to get the C
component.

\subsubsection{The power spectra}
\label{sec:powerspectra}
Theories for the origin of large-scale structure predict that
the mass distribution in the Universe is a single realization of
a statistically isotropic random field.  In other words, the
Fourier components $\tilde\delta(\veck)$ of the fractional
density perturbation $\delta(\vecx) =
[\rho(\vecx)-\bar\rho]/\bar\rho$ [where $\rho(\vecx)$ is the
density at comoving position $\vecx$ and $\bar\rho$ is the
universal mean density] are random variables that have
expectation values $\VEV{\tilde\delta(\veck)}=0$ and covariance
given by
\begin{equation}
     \VEV{\tilde \delta(\veck) \tilde \delta(\veck')} = (2\pi)^3 
     \, \delta_D(\veck+\veck') \, P_s(k).
\label{eq:scalarspectrum}
\end{equation}
Here $P_s(k)$ is the scalar power spectrum (so called because
density perturbations produce scalar perturbations to the
spacetime metric), or alternatively, the power spectrum for the
spatial mass distribution. 
Statistical isotropy demands that
the power spectrum depends only on the amplitude (rather than
orientation) of $\veck$. 

Because the temperature perturbation and polarization of the CMB
are due to density perturbations, the $a_{(lm)}^{\rm X}$
must be random variables with zero mean, $\VEV{a_{(lm)}^{\rm
X}}=0$, and covariance,
\begin{equation}
     \VEV{\left(a_{(l'm')}^{\rm X'} \right)^* a_{(lm)}^{\rm X}}
     = C_l^{{\rm XX}'} \delta_{ll'}\delta_{mm'},
\label{eq:covariance}
\end{equation}
for ${\rm X},{\rm X}' = \{{\rm T,G,C}\}$.  The statistical
independence of each $lm$ mode (i.e.\ the presence of the
Kronecker deltas) is a consequence of statistical
isotropy.  The scalar spherical harmonics $Y_{(lm)}$ and the
gradient tensor spherical harmonics $Y_{(lm)}^{\rm G}$ have
parity $(-1)^l$, whereas the curl tensor spherical harmonics
$Y_{(lm)}^{\rm C}$ have the opposite parity, $(-1)^{l+1}$.
Thus,  $C_l^{\rm TC}=C_l^{\rm GC}=0$ if the physics that gives rise
to temperature anisotropies and polarization is 
parity-invariant.  In this case, the two-point statistics of the CMB
temperature-polarization map are completely specified 
by the four sets of moments, $C_l^{\rm TT}$, $C_l^{\rm TG}$,
$C_l^{\rm GG}$, and $C_l^{\rm CC}$.  Nonzero $C_l^{\rm TC}$ or
$C_l^{\rm GC}$ would provide a signature of cosmological parity
breaking.

\subsubsection{Angular correlation functions}
The temperature two-point correlation function is
\begin{equation}
   C^{\rm TT}(\alpha) = \VEV{ {\Delta T(\hatm)\over T} {\Delta T(\hatn)
     \over T}}_{\hatm \cdot   \hatn=\cos\alpha},
\label{eq:correlationfns}
\end{equation}
where the average is over all pairs of points on the sky
separated by an angle $\alpha$.  It can be written in terms of
the temperature power spectrum (i.e.\ the $C_l^{\rm TT}$) as
\begin{equation}
     C^{\rm TT}(\alpha) = \sum_l {2 l +1 \over 4 \pi}
     C^{\rm TT}_l P_l(\cos\alpha),
\label{eq:corrpower}
\end{equation}
where $P_l(\cos\alpha)$ are Legendre polynomials.  Likewise,
orthonormality of Legendre polynomials guarantees that the 
multipole coefficients, $C_l^{\rm TT}$, can be written as
integrals over the product of the correlation function and a
Legendre polynomial.  Thus, specification of the correlation function is 
equivalent to specification of the power spectrum, and vice 
versa.  CMB theorists and experimentalists have now adopted the 
convention of showing model predictions and presenting
experimental results as power spectra ($C_l$) rather than as
correlation functions, and we subsequently stick to this
convention.  Auto- and cross-correlation
functions for the Stokes parameters and temperature-polarization
cross-correlation functions can also be defined and written in
terms of the polarization and temperature-polarization power
spectra \cite{KamKosSte97b}, but we do not list them here.

In practice, the temperature intensity (or polarization)
can never be determined 
at a given point on the sky; it can only be measured by a
receiver of some finite angular resolution (referred to as a
``finite beamwidth'').  Thus, the correlation function in
Equation~\ref{eq:corrpower} cannot be measured; one can only
measure a smoothed version.  Likewise, a finite beamwidth
$\theta_{\rm fwhm}$ (at full-width half maximum) limits
determination of the power spectrum to multipole moments
$l\la 200\,(\theta_{\rm fwhm}/1^\circ)^{-1}$.

The Differential Microwave Radiometer (DMR) experiment
\cite{Smoetal92} aboard \COBE\ produced the first map of the
temperature of the
CMB.  The receivers also provided some information on the 
polarization, but the sensitivity was not sufficient to
detect the signal expected in most cosmological models.
Measurements of the CMB intensity
were made at three different frequencies (31.5, 53, and 90 GHz)
near the blackbody peak to disentangle the possible contribution 
of foreground contaminants (e.g.\ dust or synchrotron emission)
{}from the Galaxy, as these would have a frequency spectrum that
differs from a blackbody.
The DMR beamwidth was $7^\circ$, so the temperature power
spectrum was recoverable only for $l\la 15$.  MAP, scheduled for 
launch in the year 2000, will map the sky with an angular
resolution better than $0.3^\circ$ ($l\la700$).  MAP should have
sufficient sensitivity to see the polarization, although probably not
enough to map the polarization power spectra precisely (the
polarization is expected to be roughly an order of magnitude
smaller than the temperature anisotropy).  The
Planck Surveyor, scheduled for launch around 2007, will map the
temperature and polarization with even finer angular
resolution (out to $l\la2000-3000$).  Its sensitivity should
be sufficient to map the polarization power spectra expected
{}from density perturbations (discussed below) with good precision.

\subsection{Gaussianity}

Angular three-point and higher $n$-point temperature correlation
functions can be constructed analogously to the two-point
correlation functions in Equation~\ref{eq:correlationfns}.
Fourier analogs of higher-order correlation functions can
be defined. In particular, the temperature bispectrum
$B(l_1,l_2,l_3)$ is the $l$-space version of the temperature
three-point correlation function.  It is defined by
\begin{equation}
     \VEV{a^{\rm T}_{(l_1 m_1)} a^{\rm T}_{(l_2 m_2)} a^{\rm
     T}_{(l_3 m_3)}} = \left( \begin{array}{ccc} l_1 & l_2 & l_3
     \\ m_1 & m_2 & m_3 \\ \end{array} \right) B(l_1,l_2,l_3),
\label{eq:bispectrum}
\end{equation}
where the array is the Wigner $3j$ symbol. 
This particular $m$ and $l$ dependence follows from
the assumption of statistical isotropy.  Closely related
statistics include the skewness and kurtosis (respectively, the
three- and four-point correlation functions at zero lag)
\cite{ScaVit91,LuoSch93,Luo94} and higher cumulants \cite{FerMagSil97}.
As discussed further below,
inflationary models predict the primordial distribution of
perturbations to be perfectly (or very nearly) Gaussian.
Gaussianity dictates that all the odd-$n$ $n$-point correlation
functions vanish and that for even $n$, the higher $n$-point correlation
functions can be given in terms of the two-point correlation
function.

Numerous other statistical tests of CMB Gaussianity have also
been proposed, including (but not limited to) topology of
temperature contours \cite{Col88,Gotetal90,Kogetal96} 
and the related Minkowski
functionals \cite{WinKos97,SchGor98}, temperature peak
statistics \cite{BonEfs87,VitJus87,Kogetal96}, Fourier space patterns
\cite{FerMag97,LewAlbMag99}, 
and wavelet analysis \cite{PanValFan98,HobJonLas98}.

\begin{figure}
\centerline{\psfig{file=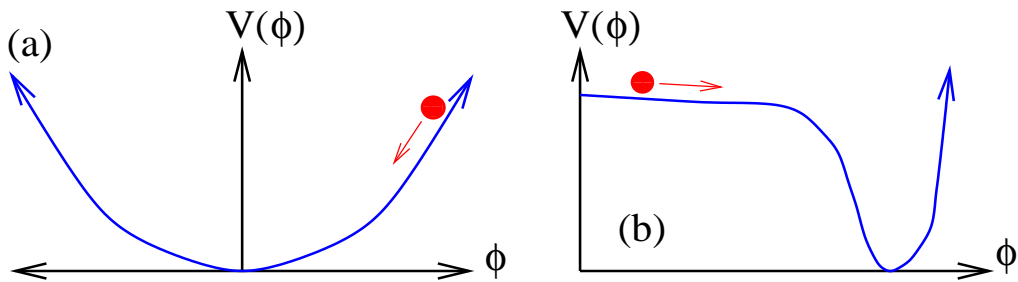,width=32pc}}
\caption{Two toy models for the inflationary potential.}
\vskip-12pt
\label{fig:potentials}
\end{figure}

\section{Predictions of Inflation}
\label{sec:inflationpredictions}

If the energy density of the Universe is dominated by matter or
radiation, then the expansion of the Universe is decelerating.
If so, the horizon grows more rapidly than the scale
factor.  In such a Universe, objects that are now beyond our
horizon and therefore inaccessible to us will eventually enter
the horizon and become visible.  Thus, the observable Universe
contains more information and is more complicated at later
times.  Inflation postulates the existence of a period of
accelerated expansion in the early Universe.  In such a
Universe, the scale factor grows more rapidly than the horizon.
Thus, objects currently visible to any given observer will
eventually exit that observer's horizon (in much the same
way that objects that fall into a black hole disappear when they
pass through the black hole's event horizon).  A period of
accelerated expansion therefore makes the Universe simpler and
smoother.
As we now discuss, this accelerated expansion also generically
drives the observable Universe to be flat and provides a
mechanism for producing primordial density perturbations and
gravitational waves.

\subsection{Scalar-Field Dynamics}
\label{sec:scalarfields}

Inflation supposes the existence of some new scalar field $\phi$ (the
``inflaton''), with a
potential $V(\phi)$ that 
roughly resembles either of those
shown in Figure \ref{fig:potentials}.  The shape is not particularly
important.  All we require is that, at some time in the
history of the Universe, the field is displaced from the minimum
of the potential, and then it rolls slowly
How slowly is slowly enough?  This is determined by the Friedmann
equation,
\begin{equation}
     H^2 \equiv \left( {\dot a \over a}\right)^2 ={8 \pi G\rho
     \over 3} - {k\over a^2} = {8 \pi G
     \over 3} \left({1\over2} \dot\phi^2 + V(\phi) \right) -{k
     \over a^2},
\label{eq:friedmann}
\end{equation}
which governs the time $t$ dependence of the scale factor
$a(t)$ of the Universe (the dot denotes derivative with
respect to time), as well as the scalar-field equation of motion,
\begin{equation}
     \ddot \phi + 3H \dot\phi +V'(\phi)=0,
\label{eq:eom}
\end{equation}
where $V'=dV/d\phi$.
In Equation~\ref{eq:friedmann}, $\rho$ is the energy density of the
Universe, which is assumed to be dominated by the inflaton
potential-energy density $V(\phi)$ and kinetic-energy density
$\dot\phi^2/2$.  The term
$k/a^2$ is the curvature term, and $k>0$, $k<0$, or $k=0$ for a
closed, open, or flat Universe, respectively.  Note that the
expansion acts as a friction term for the scalar-field equation
of motion in Equation~\ref{eq:eom}.  If
\begin{equation}
     \epsilon\equiv {m_{\rm Pl}^2 \over 16 \pi} \left({ V' \over 
     V} \right)^2 \ll 1,
\label{eq:epsilon}
\end{equation}
and
\begin{equation}
     \eta \equiv {m_{\rm Pl}^2 \over 8\pi} \left[{ V'' \over V}
     - {1 \over 2} \left({V' \over V} \right)^2 \right] \ll 1,
\label{eq:eta}
\end{equation}
where $m_{\rm Pl}=1.22 \times 10^{19}$ GeV is the Planck mass,
then the field rolls slowly enough so that the requirement for
acceleration [$p<-\rho/3$, where $p= (1/2)\dot\phi^2-V$ is the
pressure and $\rho= (1/2)\dot\phi^2 +V$ is the energy
density] is satisfied.  (Note that the definitions of $\epsilon$ and
especially of $\eta$ may differ in some papers.)

The identity of the inflaton remains a mystery.  It was originally
hypothesized to be associated with a Higgs field in
grand unified theories, 
but it may also have something to do
with Peccei-Quinn symmetry breaking, a dilaton field,
electroweak-symmetry breaking \cite{KnoTur93}, some new
pseudo-Nambu-Goldstone symmetry \cite{FreFriOli90,Adaetal93},
supersymmetry \cite{RanSolGut96}, or some other new physics.  As
discussed below, the primary predictions of slow-roll inflation
do not depend on the details of the physics responsible for
inflation but rather on some gross features that are easily
quantified.

\subsection{The Geometry}

Given any inflationary potential $V(\phi)$, the equations of
motion in Equations \ref{eq:eom} and \ref{eq:epsilon} can be
solved numerically, if not analytically.
Heuristically, during inflation, the potential $V(\phi)$ is
roughly constant, and $\dot\phi^2 \ll V(\phi)$.  If the curvature term is
appreciable initially, it rapidly decays relative to the
potential term as the Universe expands, and the solution for the 
scale factor approaches an exponential, $a(t) \propto
e^{-Ht}$.  If $k$ is nonzero initially, the curvature term is
then driven exponentially to zero during the inflationary epoch.  
In other words, if the duration of inflation is sufficiently
long to place the observable Universe in a causally connected
pre-inflationary patch, then the curvature radius is generically
inflated to an exponentially (and unobservably) large value.  In 
the language above, any initial nonzero curvature disappears
beyond the horizon during accelerated expansion.
Thus, the first prediction of slow-roll inflation is that
the Universe should be flat today; i.e.\ the total density of
all components of matter should sum to the critical density.  

\subsubsection{``Open inflation''}  
It is, of course, mathematically possible that
inflation did occur but that the inflationary epoch was 
prematurely
terminated \cite{LytSte90,KamSpe94} at just the right time so that the
Universe today would be open with density
$\Omega_0\simeq0.3$.  Such a model requires some additional
mechanism (e.g.\ another prior period of inflation and/or some
arbitrary new ``feature'' in the inflaton potential) to solve the
isotropy problem as well as to produce density perturbations.  
Several such open-inflation models have recently been constructed
\cite{Got82,LytSte90,BucGolTur95,RatPee95,Lin95,CorSpeSta96,HawTur98},
motivated by observations that suggest $\Omega_0\simeq0.3$.
The predictions of a scale-invariant spectrum and Gaussian
perturbations (discussed below) are the same as in
ordinary inflation, but the Universe would be open.
We do not find these models even nearly as compelling as
the ordinary slow-roll
models that produce a flat Universe, although some theorists may
disagree.  Fortunately, the correct model will not be determined by
debate; forthcoming CMB measurements, described below,
should distinguish conclusively between these two classes---simple and
elegant versus complicated and unappealing---of inflationary models.

\subsection{Density Perturbations, Gravitational Waves, and the
Inflationary Observables}

\subsubsection{Production of density perturbations}
Density perturbations are produced as a result of novel
quantum-mechanical effects (analogous to the production of
Hawking radiation from black holes) that occur in a Universe
with accelerating expansion 
\cite{GutPi82,Haw82,Lin82b,Sta82,BarSteTur83}.  
This process has been reviewed in detail recently \cite{Lin90,Lidetal97}, 
so here we review the physics only heuristically.
Consider perturbations $\delta\phi(\vecx,t)$ (as a
function of comoving position $\vecx$) to the
homogeneous slowly rolling field $\phi(t)$.  These perturbations
satisfy a massless Klein-Gordon equation, and the equation
of motion for each Fourier mode $\widetilde{\delta\phi}(\veck)$
is that of a simple harmonic oscillator in an expanding Universe.
At sufficiently early times, when the wavelength of any given
Fourier mode is less than the Hubble radius $H^{-1}$, it undergoes
quantum-mechanical zero-point oscillations.  
However, if the expansion is accelerating, then the
physical wavelength of this comoving scale grows faster than
the Hubble radius and eventually becomes larger than $H^{-1}$.  At this point, 
crests and troughs of a given Fourier mode can no longer
communicate, and the zero-point
fluctuation becomes frozen in as a classical perturbation
$\delta\phi(\vecx)$ to the scalar field.  Because the inflaton
potential is not perfectly flat, this induces perturbations to
the density $\delta\rho(\vecx) =(\partial V/\partial
\phi)\delta\phi(\vecx)$.

\subsubsection{Production of gravitational waves}
Tensor perturbations to the spacetime metric (i.e.\
gravitational waves) satisfy a massless Klein-Gordon equation.
A stochastic background of gravitational waves are
therefore produced in the same way as classical perturbations to 
the inflaton are produced \cite{AbbWis84}.
Moreover, the power spectra for the
inflaton-field perturbations and for the tensor metric
perturbations should be 
identical.  The power spectrum of density perturbations is a little
different from that for gravitational waves because a density
perturbation is related to a
scalar-field perturbation by $\delta\rho = (\partial
V/\partial\phi)\delta\phi$.  The production of scalar and tensor
perturbations depends only on the expansion rate during
inflation.  If the expansion rate were perfectly constant during 
inflation, it would produce flat scalar and tensor power
spectra, $P_s \propto k$ (the ``Peebles-Harrison-Zeldovich''
\cite{PeeYu70,Har70,Zel72} spectrum) and $P_t(k) \propto {\rm
constant}$.

\subsubsection{Inflationary observables}
A constant expansion rate is an oversimplification
because the field must in fact be rolling slowly down the
potential during inflation.  Given any specific functional form
for the potential, it is straightforward, using the tools of
quantum field theory in curved spacetimes (see e.g.\ 
\cite{BirDav82}), to predict precisely the functional forms
of $P_s(k)$ and $P_t(k)$.  Measurement of these power spectra
could then be used to reconstruct the inflaton potential
\cite{Lidetal97}.  Since the field must be
rolling fairly slowly during inflation, a good approximation (in 
most models) can be obtained by expanding about a constant
expansion rate.  In this slow-roll approximation, the primordial
scalar power spectrum is
\footnote{Note that this is the
spectrum for the primordial perturbations.  After the Universe
becomes matter dominated 
at a redshift $z\simeq10^4$, density
perturbations grow via
gravitational infall, and the growth factor depends on the
wave number.  Therefore, the power spectrum for matter today is
different from the primordial spectrum (it becomes $k^{-4}$ times
the primordial spectrum at large $k$), but it is straightforward 
to relate the primordial and current power spectra.}
\begin{equation}
     P_s = A_s k^{n_s},
\label{eq:scalarPs}
\end{equation}
and the primordial power spectrum for tensor perturbations is
\begin{equation}
     P_t = A_t k^{n_t}.
\label{eq:tensorPt}
\end{equation}
The amplitudes $A_t$ and $A_s$ and power-law indices $n_s$ and
$n_t$ have come to be known as the ``inflationary
observables.''    These parameters can
provide information on the inflaton potential.  In the slow-roll
approximation, the power-spectrum indices are roughly constant and given by
\cite{Sta85,LidLyt92,Davetal92,LucMatMol92,LidCol92,Tur93a,Adaetal93,Tur93b,Copetal93a,Copetal93b,Copetal94,LidTur94,Lidetal97}
\begin{equation}
     n_s=1-4\epsilon+2\eta,    \qquad  n_t= -2\epsilon,
\label{eq:nsnt}
\end{equation}
where $\epsilon$ and $\eta$ are the slow-roll parameters given
in Equations \ref{eq:epsilon} and \ref{eq:eta}.  Strictly
speaking, $\epsilon$ and $\eta$ may change (logarithmically with 
$k$) during inflation \cite{KosTur95,Lidetal97}, but, as the name
implies, the field rolls slowly during slow-roll
inflation, so the running of the spectral indices is, for all
practical purposes, very small.

The amplitudes $A_s$ and $A_t$ are similarly fixed by the
inflaton potential, but their precise values depend on
Fourier conventions and on how the scale factor today is
chosen.  However, $A_s$ and $A_t$ are proportional, respectively, to the
amplitude of the scalar and tensor contributions
to $C_2^{\rm TT}$, the quadrupole moment of the CMB temperature.
In terms of the slow-roll parameter $\epsilon$ and height $V$ of the
inflaton potential during inflation, these CMB observables are
\begin{eqnarray}
     {\cal S} &\equiv & 6\, C_2^{{\rm TT},{\rm scalar}}=
     0.66\, {V \over \epsilon m_{\rm Pl}^4}
          \nonumber\\
     {\cal T} &\equiv & 6\, C_2^{{\rm TT},{\rm tensor}}= 9.2 {V
     \over m_{\rm Pl}^4}.
\label{eq:amplitudes}
\end{eqnarray}
For nearly scale-invariant spectra, \COBE\ fixes $C_2^{\rm
TT}=C_2^{\rm TT,scalar}+C_2^{\rm
TT,tensor}=(1.0\pm0.1)\times10^{-10}$.
In terms of the slow-roll parameters, the tensor-to-scalar ratio 
is usually defined to be
\begin{equation}
     r \equiv { {\cal T} \over {\cal S}} = 13.7\,\epsilon.
\label{eq:TtoS}
\end{equation}
Comparing Equation~\ref{eq:TtoS} with Equation~\ref{eq:nsnt}, we 
observer that the observables
$n_t$ and $r$ must satisfy a consistency relation, $n_t=-0.145
r$, in slow-roll models.

To summarize, slow-roll inflation models (which account for the
overwhelming majority of inflation models that appear in the
literature) are parameterized by (\textit{a}) the height $V$ of the
inflaton potential (i.e.\ the energy scale of inflation), (\textit{b})
$\epsilon$, which depends on the first derivative $V'$ of the
inflaton potential, and (\textit{c}) $\eta$, which depends additionally on 
the second derivative $V''$.  

The discussion above suggests that because the inflaton is always
rolling down the potential, the scalar spectral index must be
$n_s<1$.  Although this may be true for simple single-field
inflation models, more complicated models (e.g.\ with multiple
fields or with different potentials) may produce ``blue''
spectra with $n_s>1$ \cite{MolMatLuc94}.

\COBE\ alone already constrains $V^{1/4} \leq 2.3\times10^{16}$
GeV.  With some additional (but reasonable) modeling, the \COBE\
constraint can be combined with current degree-scale
CMB-anisotropy measurements and large-scale-structure
observations to reduce this to $V^{1/4} \leq 1.7\times10^{16}$
GeV (e.g.\ \cite{ZibScoWhi99}).
The \COBE\ anisotropy implies $n_s=1.1\pm0.3$ 
if it is attributed entirely to density perturbations
\cite{Benetal96}, or $n_t=0.2\pm0.3$ if it is
attributed entirely to gravitational waves.  Therefore, barring
strange coincidences, the \COBE\ spectral index and relations
above seem to suggest that if slow-roll inflation is right, then
the scalar and tensor spectra must both be nearly scale-invariant
($n_s\simeq 1$ and $n_t\simeq 0$).

\subsection{Character of Primordial Perturbations}

\subsubsection{Adiabatic versus isocurvature}
The density perturbations
produced by quantum fluctuations in the inflaton field are
referred to as adiabatic, curvature, or
isentropic perturbations.  
These are perturbations to the
total density of the Universe, or equivalently, scalar
perturbations to the spacetime metric. Adiabaticity further
implies that the spatial distribution of each species in the
Universe (e.g.\ baryons, photons, neutrinos, dark matter) is the
same---that is, the ratio of number densities of any two of
these species is everywhere the same.

Adiabatic perturbations can be contrasted with primordial
isocurvature,  or equivalently, pressure or entropy
perturbations, perturbations to the ratios between the various
species in the Universe (usually in a Universe with a
homogeneous total density).  Such varying ratios would set up
perturbations to the pressure or equivalently to the entropy.
When two initially causally-disconnected regions with different
pressures come into causal contact, the pressure perturbations
push matter around, thus seeding large-scale structure. 

{\it Axion Inflation}\\
Although adiabatic perturbations are generically produced
during inflation, it is also possible to obtain isocurvature
perturbations. One example is
isocurvature perturbations to an axion density from quantum
fluctuations in the Peccei-Quinn field during inflation
\cite{SecTur85,TurWil91,AxeBraTur83,SteTur83,KofLin87,Lyt90,Lin91}.
As discussed below, comparison of the measured amplitude of
CMB anisotropies with the amplitude of galaxy clustering
essentially rules out these models. Inflation models
that produce both adiabatic and isocurvature perturbations
have also been considered
\cite{MukZel91,PolSta92,PetPolSta94,StaYok95,GarWan96,SasSte96,KawSugYan96,LinMuk97,MukSte98,Kanetal98a,Kanetal98b};
future experiments should tightly constrain the relative
contributions of the two types of perturbations.

\subsubsection{Causal versus acausal}
Perturbations produced by inflation are said to be
``super-horizon'' or ``acausal.''  This simply refers to the
fact that inflation produces a primordial (meaning
before matter-radiation equality, when gravitational
amplification of perturbations can begin) spectrum of
perturbations of all wavelengths, including those much
larger than the Hubble length at any given time.  This is to be
contrasted, for example, with ``causal'' models of structure
formation, in which perturbations are generated by
causal physics on scales smaller than the horizon.  
Since inflation implies distance scales much
larger than the Hubble length can be within a causally connected
pre-inflationary patch, the term acausal is really a
misnomer.

\subsubsection{(Nearly) Gaussian distribution of perturbations}
If the inflaton potential is flat enough for the slow-roll
approximation to be valid, then each Fourier mode of the
inflaton perturbation evolves independently; that is, the
inflaton behaves essentially like an uncoupled massless
scalar field.  As a result, inflation predicts that the
primordial density field is a realization of a Gaussian random
field: each Fourier mode is decoupled from every other, and the
probability distribution for each is Gaussian.

Of course, Gaussianity is an approximation that becomes
increasingly valid in the slow-roll limit, in which
the inflaton perturbation can be treated as a
noninteracting scalar field.
Deviations from
Gaussianity are generally accepted to be small, and most
theorists have adopted a pure Gaussian distribution as a
prediction of inflation.  However, the deviations in some models 
might be observable, and if so, would shed light on the physics
responsible for inflation
\cite{AllGriWis87,SalBonBar89,Sal92,FalRanSre93,Ganetal94,Gan94}.

This can be quantified more precisely with the three-point
statistic \cite{FerMagGor98},
\begin{equation}
     I_l^3 \equiv {1 \over (2l+1)^{3/2} (C_l^{\rm TT})^{3/2}}
     \left( \begin{array}{ccc} l & l & l
     \\ 0 & 0 & 0 \\ \end{array} \right) B(l,l,l).
\end{equation}
In slow-roll models with smooth inflaton potentials, the
prediction for this quantity is (L Wang, M Kamionkowski,
manuscript in preparation)
\begin{equation}
     \sqrt{l(l+1)}I_l^3 = {2 \over m_{\rm Pl}^2} \sqrt{ 3V
     \over \epsilon} (3\epsilon -2\eta).  
\label{eq:Il3prediction}
\end{equation}
Thus, in slow-roll models, one expects $I_0 \la 10^{-6}$ (unless 
for some unforeseen reason $\epsilon$ is extremely small and
$\eta$ is not), too small to be observed.  A larger non-Gaussian
signal may conceivably arise if there is a glitch in the
inflaton potential, but even this non-Gaussianity would be
extremely small (L Wang, M Kamionkowski, manuscript in
preparation).  Detection of nonzero $I_0$ would thus rule out
the simplest slow-roll models.

Note that the theory predicts that the
\textit{primordial} distribution of perturbations is Gaussian.
When the Universe becomes matter dominated, and density
perturbations undergo gravitational amplification, an initially
Gaussian distribution will become non-Gaussian \cite{Pee80}.
Such departures from initial Gaussianity have a specific
form and may be probed as consistency checks of inflation with
galaxy surveys that probe the matter distribution today.

\subsection{Brief Overview of Models}

A huge literature is devoted to construction of
inflationary models (for a comprehensive review, see 
\cite{LytRio99}).
Here we follow the classification of Dodelson
et al \cite{DodKinKol97}.  Models can be regarded as either large-field,
small-field, or hybrid models.  Linear models live
at the border of large- and small-field models.  In large-field
(small-field) models, the inflaton moves a distance $\Delta\phi$ that is
large (small) compared with the Planck mass during inflation.
Hybrid models introduce a second scalar field and allow a
broader range of phenomenology.

\begin{figure}
\centerline{\psfig{file=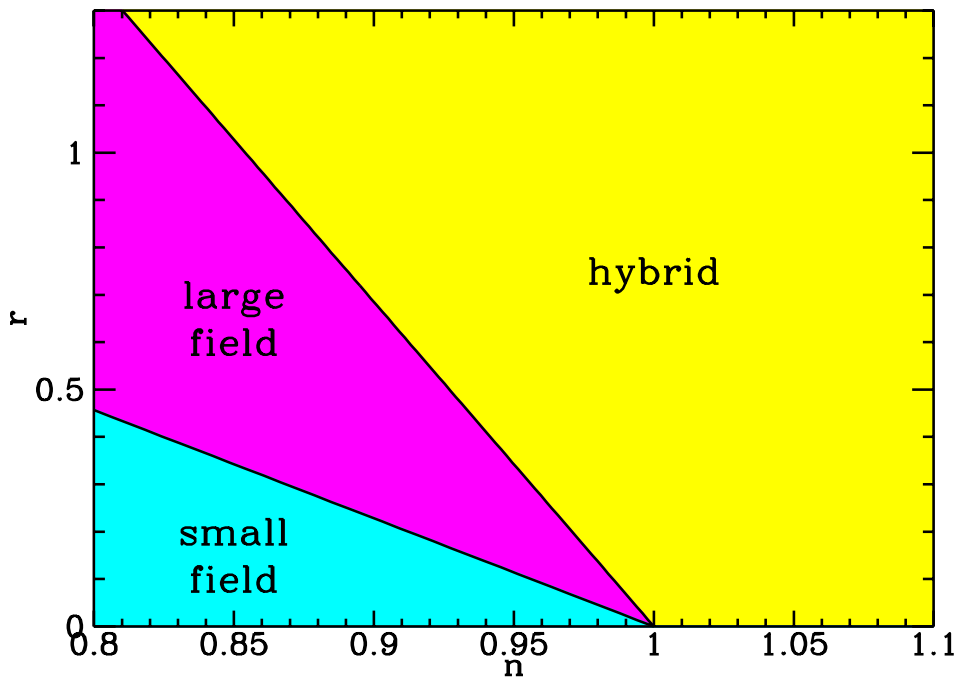,width=32pc}}
\caption{Regions in the $n_s$-$r$ plane occupied by the various
     classes of inflationary models.  (From References
     \cite{DodKinKol97,Kin98};  their $n$ is our $n_s$.)} 
\vskip-12pt
\label{fig:regions}
\end{figure}

The models can be distinguished experimentally by the values of
$V$, $\epsilon$, and $\eta$ that they predict, or equivalently by
the set of $V$, $r$, and $n_s$, as shown in
Figure \ref{fig:regions}. 
Examples of large-field models are single-field models with
polynomial potentials, $V(\phi) \propto (\phi/\mu)^p$ (with
$p>1$), or in the $p\rightarrow \infty$ limit, exponential potentials, $V(\phi)
\propto \exp(\phi/\mu)$.\footnote{Exponential potentials are
sometimes referred to as ``power-law
inflation,'' since the scale factor grows as a power law during
inflation in these models.}  The potentials in these models
resemble qualitatively the potential shown in Figure \ref{fig:potentials}\textit{a}.
These models have $V''>0$ and predict $\epsilon=[p/(p-2)]\eta>0$  and
$r\simeq 7 [p/(p+2)](1-n_s)$.  Thus, a large tensor amplitude is
expected for large $p$ (and therefore for exponential potentials
as well) and for a sufficiently large deviation of $n_s$ from
unity.

Figure \ref{fig:potentials}\textit{b} shows a potential typical of a
small-field model.  These are the types of potentials that often 
occur in spontaneous symmetry breaking and can be approximated
by $V(\phi) \propto [1-(\phi/\mu)^p]$.  These models have $V''<0$.
Demanding that the field 
move a distance that is small compared with $m_{\rm Pl}$ requires that
$(\phi/\mu) \ll 1$, and in this limit,
$\epsilon=[p/2(p-1)]|\eta| (\phi/\mu)^p \ll \eta$, $\eta<0$, and
$r\simeq7(1-n_s) \epsilon/|\eta|$.  Note that the slow-roll condition 
$\phi \ll \mu$ implies that $\epsilon \ll 1$, so $\epsilon \ll
\eta$.  It thus follows that $n_s \simeq 1+2\eta$, and that the
tensor amplitude in these models is expected to be very
small.  Note that both small- and large-field models predict
$n_s<1$.

Linear models live at the border of small- and large-field
models.  They have potentials $V(\phi) \propto \phi$ (i.e.\ they 
have $V''=0$) and predict $\epsilon\simeq-\eta>0$ and
$r\simeq(7/3)(1-n_s)$.

Although hybrid models generally involve multiple scalar fields, 
they can be parameterized by a single-field model with a
potential $V\propto [1+(\phi/\mu)^p]$.  These models have
$\epsilon>0$ and
\begin{equation}
     {\eta \over \epsilon} = {2 (p-1) \over p} \left({\phi
     \over\mu} \right)^{-p} \left[ 1 + {p-2 \over 2(p-1)} \left( 
     {\phi \over \mu}\right)^p \right] >0.
\label{eq:hybrid}
\end{equation}
Unlike small- or large-field models, hybrid models can (although 
are not required to) produce blue spectra, $n_s>1$.  Although
both $r$ and $n_t$ depend only on $\epsilon$ and are thus
related, there is no general relation between $r$ and $n_s$ in
hybrid models.  The tensor amplitude is only constrained to be
smaller than it is in exponential models.

\section{Cosmic Microwave Background Tests of Inflation}
\label{sec:cmbtests}

Photons from overdense regions
at the surface of last scatter are redder 
since they must
climb out of deeper potential wells (the Sachs-Wolfe effect
\cite{SacWol67}).  However, this is
really only one of a number of physical mechanisms that give
rise to temperature perturbations.  We have also
mentioned that if primordial perturbations are adiabatic, then
the gas in deeper potential wells is hotter, and this partially
offsets the reddening due to the depth of the potential.
Density perturbations 
induce peculiar velocities, and
these also produce temperature perturbations via Doppler shifts.
Growth of the
gravitational potential near the CMB surface of last scatter can
produce temperature anisotropies \cite{HuSug94a} [the early-time
integrated Sachs-Wolfe (ISW) effect], and so can the
growth of the gravitational potential at late times in a flat
cosmological-constant \cite{KofSta86} or open \cite{KamSpe94}
Universe (the late-time ISW effect).

Modern calculations of the CMB power spectra (the $C_l$) take
into account all of these effects.  The cosmological
perturbation theory underlying these calculations has been reviewed 
\cite{Bar80,KodSas84,MukFelBra92}, and solution of the
Boltzmann equations for the observed angular distribution of CMB
photons is discussed elsewhere \cite{MaBer95,Huetal95,SelZal96}.
Such calculations for the CMB power spectra from density
perturbations were developed in a series of pioneering papers
{}from 1970 until the late 1980s
\cite{PeeYu70,WilSil80,WilSil81,SilWil81,BonEfs84,VitSil84,
BonEfs87,Hol89}, 
and these calculations have been refined extensively in the post-\COBE\ 
era.  Similar calculations can also be carried out for the CMB
power spectra from gravitational waves
\cite{AbbWis84,MilVal86,HarZal93,CriDavSte93,FrePolCol94,NgNg95,Kos96,
KamKosSte97b,ZalSel97,HuWhi97}.

The calculations for both scalar and tensor power spectra
require solution of a series of several
thousand coupled differential equations for the perturbations to 
the spacetime metric, densities and velocities of baryons and
cold dark matter, and the moments of the
CMB photon and neutrino distributions.  A code for carrying out
these calculations (that required several hours for each model)
was made publicly available \cite{Ber95}. Hu \& Sugiyama 
\cite{HuSug94} came up with useful semianalytic fits to the
numerical calculations that provided some physical intuition
into the numerical results.  More recently, Seljak \& Zaldarriaga 
\cite{SelZal96,SelZal97} developed a line-of-sight
approach that speeded up the numerical calculations by several orders of
magnitude.  A code (CMBFAST) was made publicly available and has 
become widely used.

\begin{figure}
\centerline{\psfig{file=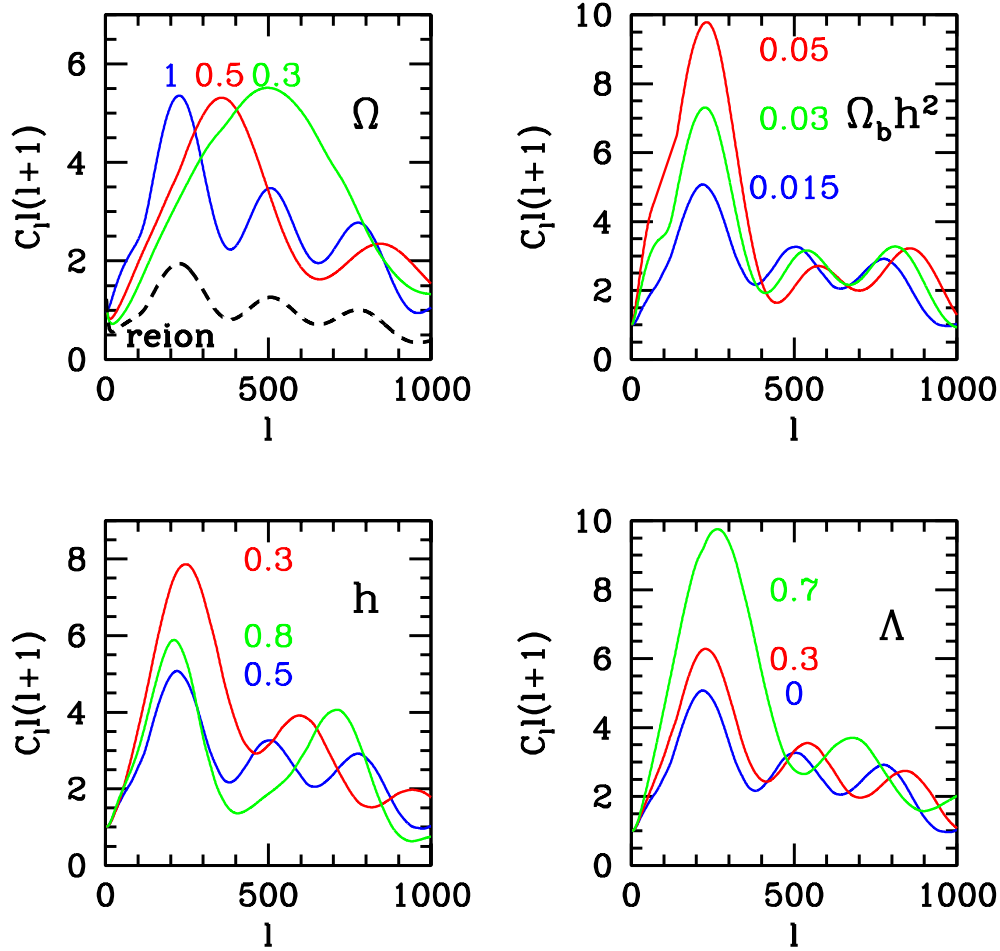,width=32pc}}
\caption{Theoretical predictions for cosmic-microwave-background
          temperature angular power spectra as a function 
	  of multipole moment $l$ for models with primordial
	  adiabatic perturbations.  
          Each graph shows the effect
	  of variation of one of these parameters.  In the lower 
	  right panel,
	  $\Omega\equiv\Omega_0+\Omega_\Lambda=1$. (From
	  Reference \cite{Junetal96b}.)}
\vskip-12pt
\label{fig:models}
\end{figure}

Given the values of the classical cosmological parameters (e.g.\ 
the nonrelativistic matter density $\Omega_0$, cosmological
constant $\Omega_\Lambda$, and baryon density $\Omega_b$, all in
units of the critical density, and the Hubble parameter $h$ in units of
100~km~sec$^{-1}$~Mpc$^{-1}$), and primordial scalar and tensor
power spectra, $P_s(k)$ and $P_t(k)$, it is straightforward to
calculate the $C_l$ with the machinery described above.
Figure~\ref{fig:models} shows results of such calculations for
models with a Peebles-Harrison-Zeldovich (i.e.\ $n_s=1$) power
spectrum of primordial adiabatic perturbations.  Each panel
shows the effect of independent variation of one of the
cosmological parameters.  As illustrated, the height, width, and
spacing of the acoustic peaks in the angular spectrum depend on
these (and other) cosmological parameters.  

The
wiggles\footnote{These are sometimes referred to inaccurately
as ``Doppler'' peaks, but are more accurately referred
to as acoustic peaks. They are sometimes called Sakharov
oscillations in honor of the scientist who first postulated the
existence of photon-baryon oscillations in the primordial plasma
\cite{Sak65}.  The existence of these peaks in the CMB power
spectrum was, to the best of our knowledge, first identified by
Sunyaev \& Zeldovich \cite{SunZel70} and Peebles \& Yu
\cite{PeeYu70}.}
come from oscillations in the photon-baryon fluid at
the surface of last scatter.  Consider an individual Fourier
mode of an initial adiabatic density perturbation.  
Because the density perturbation is nonzero initially, this
mode begins at its maximum amplitude.  The amplitude remains
fixed initially when the wavelength of the mode is larger than
the Hubble radius.  When the Universe has expanded enough that
the Hubble radius becomes larger than the wavelength of this
particular mode, then causal physics can act, and the amplitude
of this Fourier mode begins to oscillate as a standing acoustic
wave \cite{Sak65}.  Since modes with smaller wavelengths come
within the horizon earlier and oscillate more rapidly,
they have at any given time undergone more oscillations 
than longer-wavelength modes have.  The peaks evident in
Figure \ref{fig:models} arise because modes of different
wavelength are at different points of their oscillation cycles
\cite{SunZel70}.  The first peak corresponds to the mode that has had
just enough time to come within the horizon and compress
once.  The second peak corresponds to the mode that is at its maximum
amplitude after the first compression, and so forth.

\subsection{Determination of the Geometry}

These acoustic peaks in the CMB temperature power spectrum can
be used to determine the
geometry of the Universe \cite{KamSpeSug94}.  The angle
subtended by the horizon at the surface of last scatter is
$\theta_H \sim \Omega^{1/2} \;1^\circ$, where
$\Omega=\Omega_0+\Omega_\Lambda$ is the total density (objects
appear to be larger in a closed Universe than they would in a
flat Universe, and smaller in an open Universe than they would
in an flat Universe).
Moreover, the peaks in the CMB
spectrum are due to causal processes at the surface of last
scatter.  Therefore, the angles (or values of $l$) at which the
peaks occur determine the geometry of the Universe.  This is
illustrated in the top left panel of Figure~\ref{fig:models}, where the CMB spectra
for several values of $\Omega$ are shown.  As illustrated in the
other panels, the angular position of the first peak is
relatively insensitive to the values of other undetermined (or
still imprecisely determined) cosmological parameters such as
the baryon density, the Hubble constant, and the cosmological
constant.  

Small changes to the spectral index $n_s$ tilt the entire
spectrum slightly to smaller (larger) $l$ for $n_s<1$ ($n_s>1$),
and the location of the first peak is only weakly affected.  
Gravitational waves would only
add to the temperature power spectrum at $l\ll 200$ (as
discussed below in Section 4.4).  Therefore,
although gravitational waves could affect the height of the
peaks relative to the normalization at small $l$, the locations
would not be affected.  It is plausible that an early generation 
of star formation released a sufficient flux of ionizing
radiation to at least partially reionize the Universe, and if
so, these ionized electrons would rescatter some fraction
$\tau$ of the CMB photons.   A variety of theoretical arguments
suggest that a fraction $\tau = {\cal O}(0.1)$ of CMB photons
were rescattered \cite{KamSpeSug94,TegSilBla94,HaiLoe97} (and
the amplitude of anisotropy at degree scales observed already
supports this).  Although
reionization would damp the peaks by a factor $e^{-2\tau}$, as
indicated by the curve labeled ``reion'' in the top left 
panel of Figure \ref{fig:models} (but note that the figure assumes
$\tau=1$), the location of the peaks would remain unchanged.
Therefore, if peak structure is observed in the CMB power
spectrum, determination of the location of the first peak will
provide a robust determination of the geometry of the Universe
\cite{KamSpeSug94}.

\subsubsection{Open inflation}
The most distinctive signature of an open-inflation model would be a
low-$\Omega_0$ CMB power spectrum from adiabatic perturbations such as
one of those shown in the top left panel of Figure \ref{fig:models},
in which the first peak is shifted to larger $l$.
Open inflation would also produce an increase in large-angle
anisotropy from the integrated Sachs-Wolfe effect
\cite{LytSte90,KamSpe94}, but different open-inflation
models make different predictions about the large-angle anisotropy.
Moreover, determination of the CMB power spectrum at these large
angular scales is cosmic-variance limited, so it is unlikely
that large-angle CMB anisotropies alone will be able to provide
a robust test of open-inflation models.  The ISW effect may
alternatively be identified by cross-correlation of the CMB with
some tracer of the mass density along the line of sight
\cite{CriTur96,Kam96}, such as the X-ray background
\cite{BouCriTur98,KinKam98} or possibly weak-lensing maps
\cite{ZalSel99} (as discussed further in  Section
\ref{sec:cosmologicalconstant}).

\begin{figure}
\centerline{\psfig{file=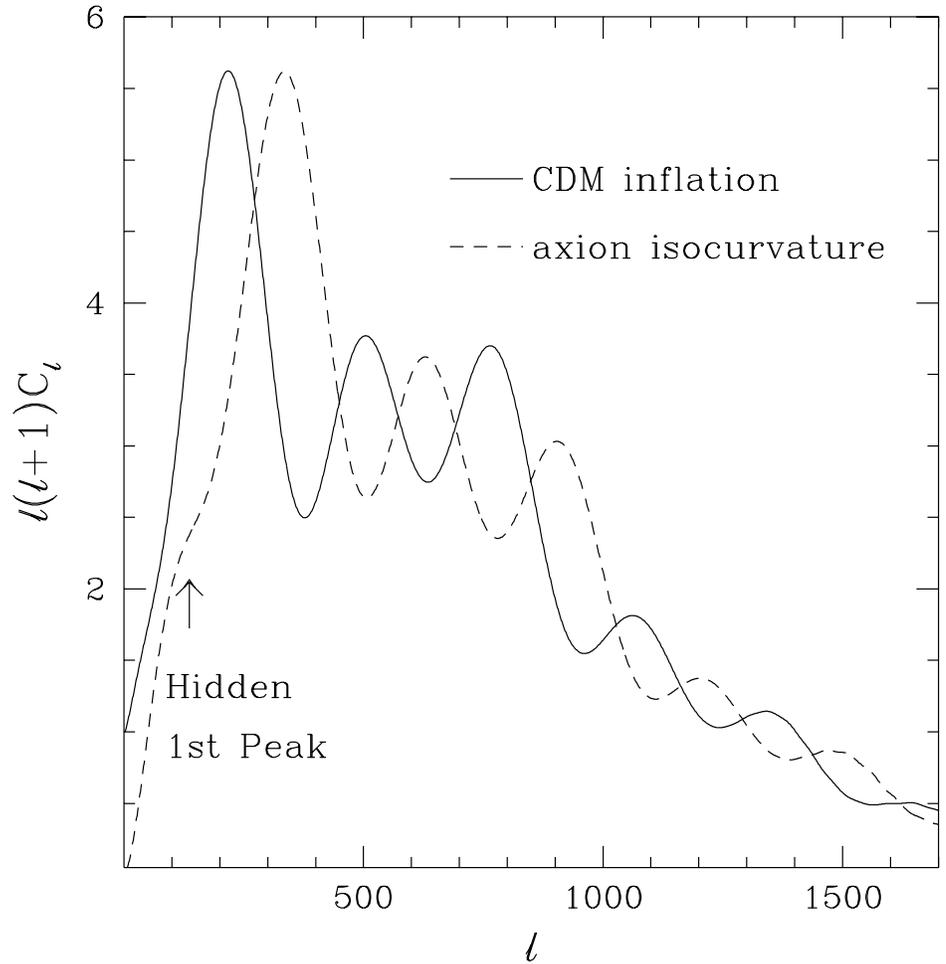,width=32pc}}
\caption{The angular power spectrum for an inflationary model with
          primordial adiabatic perturbations and for another with
	    primordial isocurvature perturbations.  
          \textit{Solid line}, cold dark matter and inflation; 
          \textit{dashed line}, axion isocurvature. (From
	  Reference \cite{HuWhi96a}.)}
\vskip-12pt
\label{fig:huwhite}
\end{figure}

\subsection{Adiabatic Versus Isocurvature Modes}

The physics described above yields a distinctive pattern in the peak
structure of the CMB power spectrum, and this leads to an important
test of inflation.  If primordial perturbations are
isocurvature rather than adiabatic,
then when a given Fourier mode comes within the
horizon and begins to oscillate, it begins to oscillate from its
minimum (rather than maximum) amplitude.  Thus, the phase of its
oscillation differs by $\pi/2$ from what it would be if the
perturbation were adiabatic.  As a result, the locations of the peaks in
the CMB power spectrum in isocurvature models differ in phase from
what they would be in adiabatic models \cite{HuSug95,HuWhi96a,Kos98}, as shown
in Figure \ref{fig:huwhite}.
It has also been shown that the
relative locations of the higher peaks differ in adiabatic and
isocurvature models, independent of the shift in the locations of the
peaks due to the geometry \cite{HuWhi96}.

\subsubsection{Axion inflation}
When the matter power spectrum is normalized to
the amplitude of galaxy clustering, isocurvature models with
nearly scale-invariant primordial power spectra (e.g.\ axion
isocurvature or ``axion inflation'' models) produce roughly six
times the CMB anisotropy seen by \COBE\ \cite{KodSas86,EfsBon87}
(since there is no cancellation between the effects of the
intrinsic temperature and the potential-well depth at the
surface of last scatter) and are thus ruled out.  

\subsection{Coherent Perturbations and Polarization}

Each Fourier component of the density field induces a Fourier
component of the peculiar-velocity field, and the oscillations
of these peculiar velocities are out of phase with the
oscillations in the density perturbation (just as the velocity
and position of a harmonic oscillator are out of phase).  These
peculiar velocities induce temperature anisotropies (via the
Doppler effect) that are thus out of phase with those from
density perturbations.  This Doppler effect fills in the troughs 
in the $C_l^{\rm TT}$, which would otherwise fall to zero.

\begin{figure}[htbp]
\centerline{\psfig{file=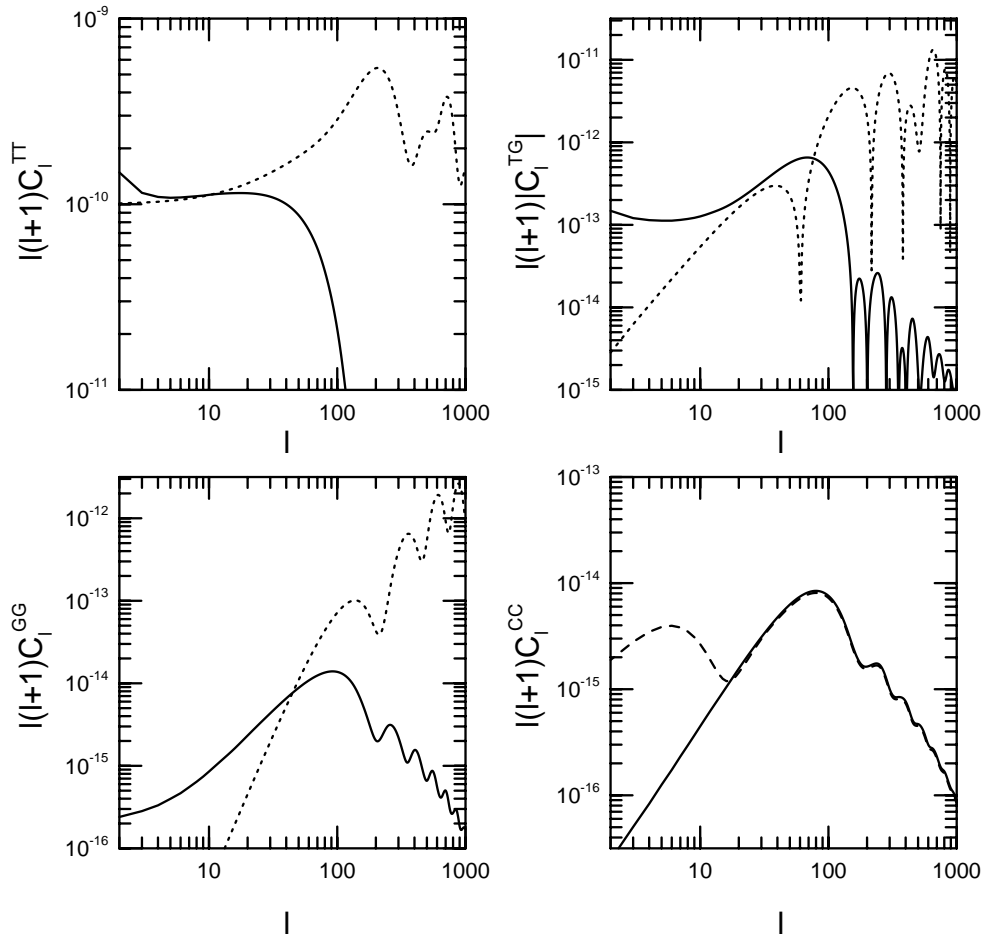,width=7in}}
\bigskip
\caption{
          Theoretical predictions for the four nonzero cosmic-microwave-background
	  temperature-polarization spectra as a function
	  of multipole moment $l$.  \textit{Solid curves} are the
	  predictions for a \COBE-normalized inflationary model
	  with no reionization and no gravitational waves for
	  $h=0.65$, $\Omega_b
	  h^2=0.024$, and $\Lambda=0$.  \textit{Dotted curves} are the
	  predictions that would be obtained if the \COBE\ 
	  anisotropy were due entirely to a stochastic
	  gravity-wave background with a flat scale-invariant
	  spectrum (with the same cosmological parameters).
	  The panel for $C_l^{\rm CC}$ 
          contains no dotted curve because scalar perturbations
	  produce no ``C'' polarization component; instead,
          the \textit{dashed line} in the \textit{bottom right panel} shows a
	  reionized model with optical depth $\tau=0.1$ to the
	  surface of last scatter.  (From Reference \cite{KamKos98}.)
       }
\label{fig:clsplot}
\end{figure}

The CMB polarization is related to
the peculiar velocity at the surface of last scatter
\cite{ZalHar95}, so the peaks in the polarization power
spectrum (from density perturbations) are out of phase from those in
the temperature power spectrum and fall close to zero 
(Figure \ref{fig:clsplot}).  This relative
positioning of the temperature and polarization peaks is a
signature of coherent perturbations (rather than those
produced, for example, by the action of topological defects, as
discussed below) \cite{Kos98}.
Zaldarriaga \& Spergel \cite{SpeZal97} argue that the location of the
first peak in the polarization power spectrum provides a test of
primordial perturbations as the origin of structure and thus of
inflation. If some causal mechanism (such as topological defects)
produced large-scale structure, the the first peak would have to occur
at smaller angular scales in order to be within the horizon at the
surface of last scatter (see below).  A peak so close to the causal
horizon could only occur with super-horizon-sized
primordial perturbations, for which inflation is the only causal 
mechanism.

\subsection{Polarization and Gravitational Waves}  

Gravitational waves are usually detected by observation of the motion
they induce in test masses.  The photon-baryon fluid at the
surface of last scatter acts as a set of test masses for
detection of gravitational waves with wavelengths comparable to
the horizon, such as those predicted by inflation.  These
motions are imprinted onto the temperature and polarization of
the CMB.  The top left panel of Figure \ref{fig:clsplot}
(\textit{solid curve}) shows the
temperature power spectrum for a \COBE-normalized flat
scale-invariant ($n_t=0$) spectrum of gravitational waves.  It
is flat and relatively featureless for $l\la70$.  
The dropoff at $l \ga70$ is due to the fact that the amplitudes
of gravitational-wave modes that enter the horizon before the
epoch of last scatter have decayed with the expansion of the
Universe.  Unfortunately, cosmic variance from scalar
perturbations provides a fundamental limit to the sensitivity of
CMB temperature maps to tensor perturbations \cite{KnoTur93}.
Even if all other cosmological parameters are somehow fixed,
a perfect temperature map can never detect an
inflaton-potential height smaller than one-tenth the upper limit 
provided by \COBE\ \cite{Junetal96b}.  More realistically, the
effects of gravitational waves and reionization on the
temperature power spectrum are similar and difficult to
disentangle, so improvements to the current \COBE\ sensitivity
to gravitational waves is unlikely with a temperature map alone.

However, with a polarization map of the CMB, the scalar and
tensor contributions to CMB polarization can be geometrically
decomposed in a model-independent fashion, and the
cosmic-variance limit present in temperature maps can thereby be
circumvented \cite{Ste96,KamKosSte97a,SelZal97}.  Scalar perturbations
have no handedness, so they cannot give rise to a curl
component.  On the other hand, tensor perturbations do
have a handedness, so they induce a curl component.  Therefore,
if any curl coefficient, $a_{(lm)}^{\rm C}$, is found to be
nonzero, it suggests the presence of gravitational
waves.\footnote{A curl component
may also be due to vector perturbations.
Although topological-defect models may excite such modes, they
do not arise in inflationary models.}

To illustrate, Figure~\ref{fig:clsplot} shows the four nonzero
temperature-polarization power spectra. The \textit{dotted curves}
correspond to a \COBE-normalized inflationary model
with no gravitational waves.  The \textit{solid
curves} show the spectra for a \COBE-normalized stochastic
gravitational-wave background.  

\subsubsection{Detectability of gravitational waves: curl
component only}
Consider a mapping experiment that measures the
polarization on the entire sky with a temperature sensitivity
$s$ (which has units $\mu$K~$\sqrt{\rm sec}$) for a time $t_{\rm
yr}$ years.  If only  the curl component of the polarization is
used to detect tensor perturbations, then such an experiment can
distinguish a tensor signal from a null result at the $2\sigma$
level if the inflaton potential height is \cite{KamKos98}
\begin{equation}
     V \ga (4\times 10^{15}\, {\rm GeV})^4 \, t_{\rm yr}^{-1} \,
     (s/\mu{\rm K}\, \sqrt{\rm sec})^2.
\label{eq:tensordetectable}
\end{equation}
Equation~\ref{eq:tensordetectable} indicates that to access an
inflaton-potential height not already excluded by \COBE\
requires a detector
sensitivity $s\la35\,t_{\rm yr}^{1/2}\,\mu$K$\sqrt{\rm
sec}$.  To compare this with realistic values, the effective
sensitivity of MAP is $s\simeq150\,t_{\rm
yr}^{1/2}\,\mu$K$\sqrt{\rm sec}$ and that for Planck is about
$s\simeq35\,t_{\rm yr}^{1/2}\,\mu$K$\sqrt{\rm sec}$, and
technological developments have improved the detector
sensitivity roughly an order of magnitude per decade for the
past several decades.  Even better sensitivities may be possible
with deep integration on a smaller region of the sky.

\subsubsection{Reionization} 
In some sense, Equation~\ref{eq:tensordetectable} is conservative 
because even a
small amount of reionization will significantly increase the
polarization signal at low $l$  (indicated by the \textit{dashed curve}
in the CC panel of Figure~\ref{fig:clsplot} \cite{Zal97}).
If $\tau=0.1$, then the sensitivity to tensor modes is increased 
by roughly a factor of five \cite{KamKos98}.

\subsubsection{Full polarization and temperature spectra}
Although searching only for the curl component provides a
model-independent probe of tensor modes, a
stochastic gravitational-wave background leads to specific
predictions for all four nonzero temperature-polarization power 
spectra (Figure \ref{fig:clsplot}).  Fitting an
inflationary model to the entire set of temperature and
polarization power spectra can improve tensor detectability,
especially at poorer sensitivities, albeit with the introduction
of some model dependence.  For detector sensitivities $s\ga
15\,t_{\rm yr}^{1/2}\,\mu$K$\sqrt{\rm sec}$, the sensitivity to
a tensor signal is improved by factors of a few or so
\cite{KamKos98}, depending on the cosmological model, whereas
for detector sensitivities $s\la 15\,t_{\rm
yr}^{1/2}\,\mu$K$\sqrt{\rm sec}$, the sensitivity is attributable
almost
entirely to the CC power spectrum and approaches the limit in
Equation~\ref{eq:tensordetectable}.

\begin{figure}[htbp]
\centerline{\psfig{file=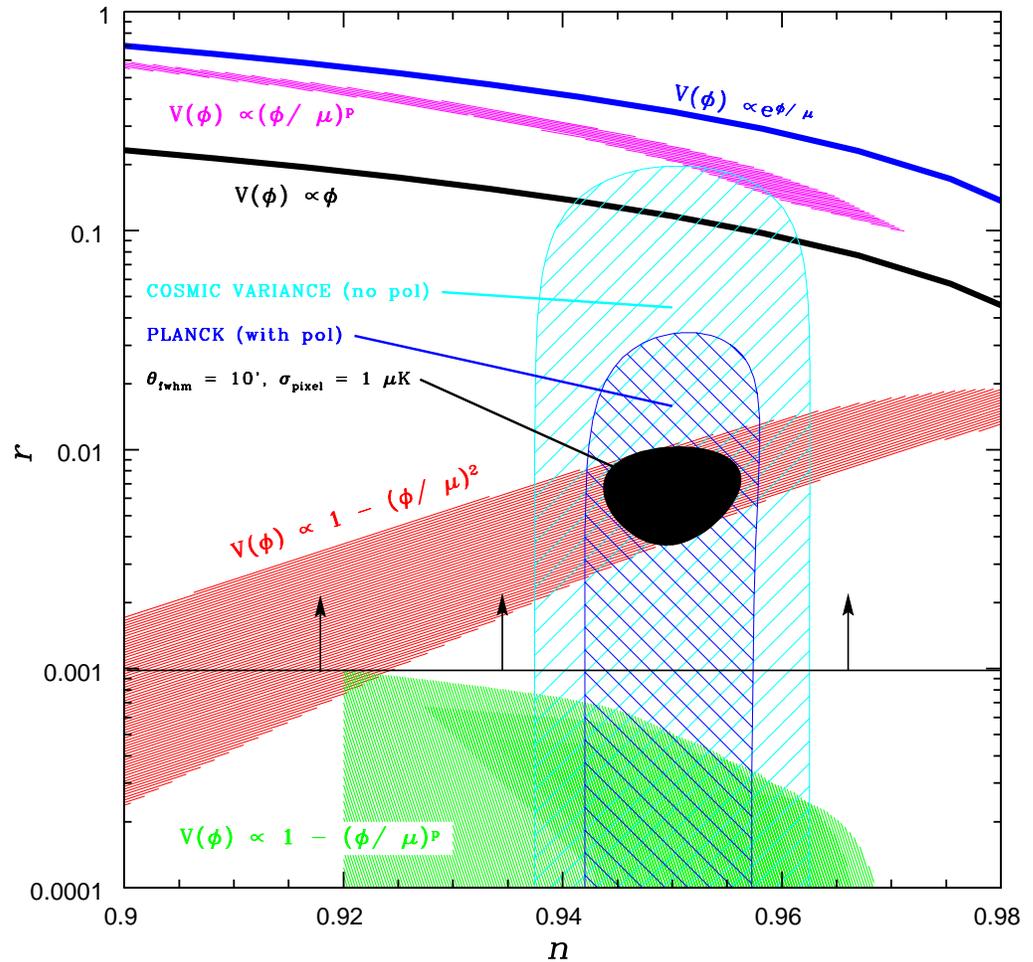,width=5.5in}}
\bigskip
\caption{Simulated $2\sigma$ error ellipses that would be
         obtained by a cosmic-variance-limited temperature map,
	 the Planck Surveyor (with polarization), and an
	 experiment with three times the sensitivity of Planck.
	 This assumes an inflationary model with $r=0.01$
	 and $n_s=0.95$ and an optical depth to the surface of
	 last scatter of $\tau=0.05$.  \textit{Shaded regions}
	 indicate the
	 predictions of various inflationary models.  \textit{Solid
	 horizontal curve} indicates the regions of this
	 logarithmic parameter space that would be accessible
	 with a putative polarization experiment with 30 times
	 the Planck instrumental sensitivity \cite{KamKos98}.
	 Even better sensitivities may be possible with deep
	 integration on a smaller region of the sky.
	 (From Reference \cite{Kin98}.)
       }
\label{fig:Plancklog}
\end{figure}

\subsubsection{Measurement of inflationary observables}
Several authors have addressed the question of
how precisely the inflationary observables can be reconstructed
in the case a positive detection of the stochastic
gravitational-wave background with only a temperature map
\cite{Kno95,Junetal96b,BonEfsTeg97,DodKinKol97} and with a
polarization map as well \cite{ZalSelSpe97,Lidetal97b,Kin98}.  
We follow the discussion of Ref. \cite{Kin98}.

Figure \ref{fig:Plancklog} shows the $2\sigma$ error ellipses
that would be obtained by the Planck Surveyor using the
temperature only (i.e. the cosmic-variance limit) and with the
polarization, assuming a gravitational-wave background with
$r\simeq0.01$ and $n_s\simeq0.9$.  (A larger gravitational-wave
amplitude would be detectable, as shown in Figures~3--6 in
Reference \cite{Kin98}.)  The $2\sigma$ cosmic-variance limit from a
temperature map is shown as is the $2\sigma$ constraint to the
parameter space expected for Planck (with polarization).
Although such a tensor
amplitude cannot be distinguished from a null result, the figure
shows (the dark shaded region) that a hypothetical experiment with three times
the Planck polarization sensitivity could discriminate between a 
such a tensor signal and a null result.  It would also
discriminate between a single small-field model and a hybrid
model.  Of course, the sensitivity to tensor modes can be
improved as the instrumental sensitivity is improved, as
indicated by Equation~\ref{eq:tensordetectable}.  For example, the thin
horizontal line at $r=0.001$ shows the smallest value of $r$
that could be distinguished from a null result by a hypothetical
one-year experiment with an instrumental sensitivity
$s=\mu$K$\sqrt{\rm sec}$, roughly 30 times that of Planck
\cite{KamKos98}.  A
null result from such an experiment would suggest that if
inflation occurred, it would have required a small-field model.

\subsection{Gaussianity}

The prediction of primordial Gaussianity or of some specific
small deviations from Gaussianity can be probed with the CMB
angular bispectrum or higher $n$-point correlation functions
discussed above.  A nonzero large-angle CMB
bispectrum may arise from the integrated Sachs-Wolfe effect if
there is a cosmological constant \cite{Ganetal94}.  Such a
bispectrum, as well as that probed by the Sunyaev-Zeldovich
effect, may be discernible via  cross-correlation between
gravitational-lensing maps and the CMB \cite{GolSpe98}. A more
powerful test of inflation models may arise from probing
the bispectrum induced by nonlinear evolution
at the surface of last scattering (S Winitzki, A Kosowsky, DN Spergel,
manuscript in preparation).

Ferreira et al and Pando et al \cite{FerMagGor98,PanValFan98}
recently claim to have already found some signature of
non-Gaussianity in the \COBE\ maps.  In particular,
Ferreira et al \cite{FerMagGor98} find $I_l^3 \sim1$ for
$l\sim16$ (although it is still not clear if the effect is real
\cite{KamJaf98}).  If this result is correct, then the simplest
slow-roll inflation models are not viable (see
Equation~\ref{eq:Il3prediction}).

\section{Topological-Defect Models}
\label{sec:defects}

The leading alternative
to structure-formation models based on inflation have been
those based on topological defects, particularly
cosmic strings \cite{Zel80,Vil81,SilVil84,Tur85,HinKib95}, 
global monopoles \cite{BarVil89,BenRhi91}, 
domain walls \cite{HilSchFry89,PreRydSpe89}, 
and textures \cite{Tur89,Tur91} 
(for reviews, see  \cite{Vil85,VilShe94}).  
Defect models postulate a phase transition
in the early Universe that leads to a vacuum manifold with
nontrivial topology; the type of defect depends
on the specific topology (see \cite{HinKib95} for a
review). Since defect formation is a process
governed by causal physics, the vacuum state of the field
must be uncorrelated on scales larger than the horizon at
the time of the phase transition, guaranteeing the
formation of defects with a characteristic length scale
of the horizon (the ``Kibble mechanism'' \cite{Kib76}).

The simplest defects are domain walls, which arise in theories
with a discrete symmetry.  Domain-wall models are not
viable because their energy densities are
large enough to produce larger CMB temperature fluctuations
than those observed \cite{ZelKozOku74,SteTur89,TurWatWid91}.

Cosmic strings are stable defects that arise in gauge models
with a $U(1)$ symmetry.  They
produce density perturbations by their gravitational
interactions with ordinary matter.
Global-monopole and texture models are
unstable defects that arise in models with a perfect 
global symmetry.  They provide two mechanisms for structure
formation: (\textit{a}) the energy-density provided by misalignment of
scalar fields as causally disconnected regions of the Universe
come into causal contact, and (\textit{b}) the explosive events that
occur when the topological defects unwind.  Non-topological
texture models \cite{Vil82,Pre80,TurSpe91,Jaf94} postulate an
even higher global symmetry and seed structure via scalar-field
alignment even though no topological defects are formed.
Generically, one expects quantum  gravity to violate global
symmetries to the level that would render global-monopole,
texture, and scalar-field-alignment models unworkable
\cite{KamMar92,Holetal92}.  If it could be shown that such 
models do seed large-scale structure, valuable information on
Planck-scale physics would thus be provided. 

\subsection{Cosmic-Microwave-Background Power Spectra in Defect Models}

In contrast to inflationary models, which lay down an initial
spectrum of density perturbations, defect models
produce perturbations actively throughout the history of the
Universe. This generally leads to loss of coherence in the
perturbations and a corresponding smoothing of the acoustic peaks
\cite{Albetal96,Magetal96a,HuWhi97c}. Defect-model perturbations are also
causal, being generated by physical processes 
inside the horizon
\cite{SteVee90,Tur96b,Tur96c,HuSpeWhi97,DurKun98}.
And finally,
primordial perturbations in defect models more closely resemble
those in primordial-isocurvature rather than those in adiabatic models
\cite{SteVee90,DurSak97,HuSpeWhi97}.
Moreover, the action of topological defects 
generically produces vector and tensor perturbations which
increase the anisotropy on small angular scales
\cite{PenSelTur97,Alletal97}, further suppressing any peak
structure (although it may produce some characteristic C
polarization \cite{SelPenTur97}).

\begin{figure}[htbp]
\centerline{\psfig{file=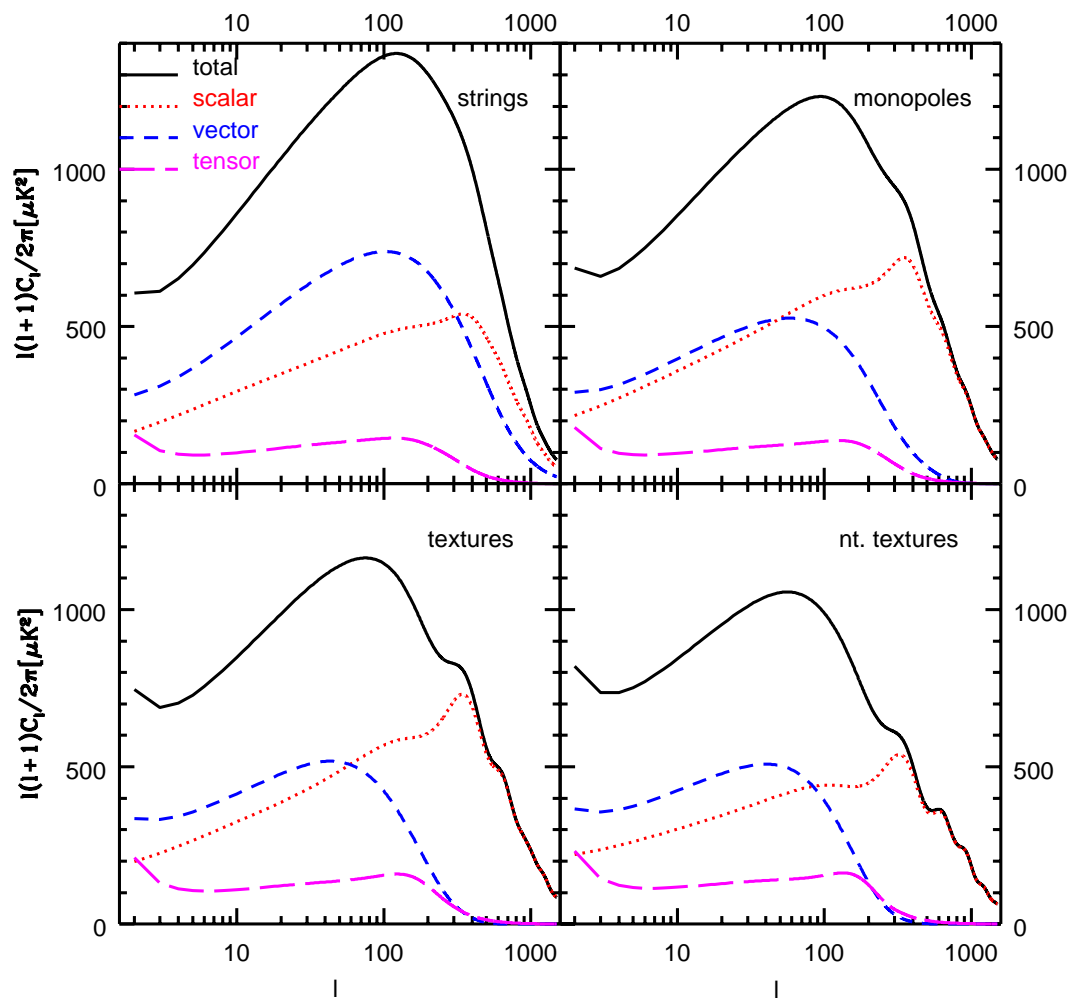,width=5.5in}}
\bigskip
\caption{Cosmic-microwave-background power spectra from topological-defect models. 
\textit{Solid line}, total; \textit{dotted}, scalar; 
\textit{short-dashed}, vector; \textit{long-dashed}, tensor. 
(From Reference \cite{SelPenTur97}.)}
\label{fig:defects}
\end{figure}

Until recently, different groups obtained
different results about the extent to which acoustic peaks
exist in defect models, and under what circumstances
\cite{CriTur95,Tur96b,DurGanSak96,DurZho96,Tur96b,Tur96c}.
Calculations of CMB power spectra based on large simulations of
a variety of defect sources have now been performed
\cite{Alletal97,PenSelTur97,ConHinMag99} (see
Figure \ref{fig:defects} for some).  The numerical results
indicate that the acoustic peaks are washed out.

At this point, it appears that the simplest defect models are
inconsistent with the observed CMB fluctuations and the large-scale
structure traced by galaxy surveys.  Although this could have
been inferred from the generic arguments discussed in the
Introduction \cite{JafSteFri94}, it has been supported by these
more recent precise calculations
\cite{AlbBatRob97,Alletal97,PenSelTur97}.  The question now is
whether any more complicated (or ``sophisticated'') defect models be 
viable.
Albrecht et al \cite{AlbBatRob98a,AlbBatRob98b} have
suggested that a cosmological constant might help improve
concordance with current data.  However, suppose the CMB
temperature power spectrum continues to look increasingly like that
caused by inflation (i.e.\ with identifiable acoustic peaks), as
new data seem to suggest.
If so, can any defect model reproduce such a power spectrum?
Turok \cite{Tur96c} produced 
a power spectrum with a phenomenological defect model that
closely mimicked an inflation power spectrum, and Hu
\cite{Hu99} has invented a similar isocurvature model.
But it is hard to see how
to position the acoustic peaks in isocurvature-like models at
the same angular scales as in adiabatic models without some
rather artificial initial conditions
\cite{HuSpeWhi97,Magetal96b}.  
It is also difficult to simultaneously account for the fluctuation
amplitude in the CMB and galaxy surveys, unless there is a
breaking of scale invariance \cite{Pen98} (possibly from some finite breaking
of the global symmetry \cite{KamMar92,Holetal92}).
It may, in fact, be possible to construct some causal
models that produce peaks in the CMB power spectrum
\cite{DurKun98,DurSak97}, but it is unclear whether
fluctuations that mimic a specific inflationary model can be
produced, particularly when additional constraints from
polarization are taken into account.
Finally, it should be noted that hybrid models with both
primordial adiabatic perturbations and defects have been
entertained \cite{Jen96,LinRio97,AveCalMar98}.

\subsection{Non-Gaussianity}

Topological-defect models may also be distinguished
by the non-Gaussian signatures they produce in the CMB.
Because the evolution of topological defects is
nonlinear, they generically produce non-Gaussian structures in
the CMB.  Put another way, the production of defects via the
Kibble mechanism is a Poisson process; the number of defects
within any volume in the Universe is Poisson distributed.  The
central-limit theorem guarantees that as the number density of
defects becomes large, the distribution should become
increasingly Gaussian.  Thus cosmic-string models should look
more like Gaussian perturbations than textures should, since the Kibble
mechanism produces roughly one texture per 25 Hubble volumes as
opposed to roughly one cosmic string per Hubble volume
\cite{SchBer91,Per93b,BenRhi93,GilPer95,SchSch95}.
In the large-$N$
limit of the $O(N)$ sigma model, the clearest signature of
non-Gaussianity from scalar-field alignment is at large angular
scales \cite{Jaf94}; on small distance scales, the theory looks roughly
Gaussian.  Constraints to non-Gaussianity from the
galaxy distribution have already posed problems for
scalar-field-alignment models for several years.  
Since defects are 
coherent structures, they can produce 
corresponding coherent structures
in the CMB temperature anisotropy.  For example, a cosmic string 
can produce a linear discontinuity in the CMB temperature
\cite{KaiSte84}, which can be searched for most efficiently
through statistics tailored to match this particular signal
\cite{MoePerBra94,Per97,Per98}. Textures might form large hot
spots \cite{TurSpe90,Tur96a}.

\section{Dark Matter}
\label{sec:darkmatter}

The CMB can potentially provide a wealth of information about the dark
matter known to dominate the mass of the Universe.
The smallness of the amplitude of CMB temperature
fluctuations has for a long time provided some of the strongest evidence
for the existence of dark matter.  In a low-density Universe, density
perturbations grow when the Universe becomes matter-dominated and end
when it becomes curvature-dominated.  If the luminous matter
($\Omega_{\rm lum} \sim10^{-3}$) were all the mass in the Universe,
then the epoch of structure formation would be too short to allow
density perturbations to grow from their early-Universe amplitude,
fixed by \COBE, to the amplitude observed today in galaxy
surveys.

More precise measurements of the CMB power spectrum hold the
promise of providing much more detailed information about the
properties and distribution of dark matter.
There are currently several very plausible dark-matter
candidates that arise from new particle physics, and  
some evidence has already been claimed for the
existence of several of these.  For example, some observational
evidence points to the existence of a cosmological constant
\cite{Peretal97,Peretal99,Rieetal98}, and the LSND experiment
suggests that  massive neutrinos may constitute a significant fraction
of the mass of the Universe \cite{Athetal95,Athetal96}.
Moreover, there are good arguments that a significant fraction
of the mass in galactic halos 
is made of some type of
cold-dark-matter particle, e.g.\ weakly interacting massive
particles (WIMPs) \cite{JunKamGri96} or axions \cite{Tur90,Raf90}.

\subsection{Cold Dark Matter}

A number
of dynamical measurements suggest that the nonrelativistic-matter
density is $\Omega_0\ga0.1$, whereas big-bang nucleosynthesis suggests a
baryon density of $\Omega_b\la0.1$.  Observations of
X-ray emission from galaxy clusters suggest that the
nonrelativistic matter in clusters outweighs the baryonic matter
by a factor of three or more \cite{Whietal93}, and weak lensing
of background galaxies by clusters directly reveals
large amounts of dark matter \cite{TysKocDel98}.  This evidence
strongly indicates the existence of some nonbaryonic dark
matter.  By fitting the power spectra from MAP and Planck to
theoretical predictions, one should simultaneously be able to
determine both $\Omega_0 h^2$ and $\Omega_b h^2$ to far better
precision than that obtained by current observations
\cite{Junetal96b,BonEfsTeg97,ZalSelSpe97}.  If a substantial
fraction of the mass in the Universe is in fact made of
nonbaryonic dark matter (e.g.\ WIMPS or axions), then
it will become evident after MAP and Planck.  Unfortunately,
there is no way to discriminate between WIMPs and axions with
the CMB.

\subsection{Neutrinos}

One of the primary goals of experimental particle
physics is pursuit of a nonzero neutrino mass.  Some
recent (still controversial) experimental results suggest that
one of the neutrinos may have a mass of $\order(5\,{\rm eV})$
\cite{Athetal95,Athetal96}.  There have been some arguments 
(again, still
controversial) that such a neutrino mass is exactly what
is required to explain apparent discrepancies between
large-scale-structure observations and the simplest
inflation-inspired standard-CDM model
\cite{ShaSte84,DalSch92,DavSumSch92,Klyetal93,Prietal95,BonPie98}.

If the neutrino does indeed have a mass of $\order(5 \, {\rm
eV})$, then roughly 30\% of the mass in the Universe is
in the form of light neutrinos.  These neutrinos will affect the
growth of gravitational-potential wells near the epoch of last
scatter, thus leaving an imprint on the CMB angular
power spectrum \cite{DodGatSte96,MaBer95,Lopetal98a}.  The
effect of a light neutrino
on the power spectrum is small, so other cosmological parameters
that might affect the shape of the power spectrum at larger
$l$ must be known well.  Eisenstein et al \cite{EisHuTeg99} argue
that by combining measurements of the CMB power spectrum with
those of the mass power spectrum measured by, for instance, the Sloan
Digital Sky Survey, a neutrino mass of $\order(5 \, {\rm eV})$
can be identified.  The CMB may
constrain the number of noninteracting relativistic
degrees of freedom in the early Universe \cite{Junetal96b}. 
Although weaker than
the bound from big-bang nucleosynthesis \cite{SteSchGun77,
CopSchTur97}, the CMB
probes a different epoch ($T\sim$eV rather than $T\sim$MeV) and
may thus be viewed as complementary.

\subsection{Cosmological Constant}
\label{sec:cosmologicalconstant}

Some recent evidence seems to point to the existence of an
accelerating expansion, possibly due to a nonzero cosmological
constant (\cite{Peretal97,Peretal99}; for a review
of the cosmological constant, see \cite{CarPreTur92}).
The CMB may help probe the existence of a cosmological constant
in a number of ways.  As discussed
above, if adiabatic perturbations are responsible for
large-scale structure, then the position of the first acoustic
peak in the CMB power spectrum
provides a model-independent probe of the total density,
$\Omega=\Omega_0 + \Omega_\Lambda$ \cite{KamSpeSug94}.  In
contrast, the supernova measurements of the Hubble diagram at
large redshifts determine primarily the deceleration
parameter $q_0 = \Omega_0/2 - \Omega_\Lambda$, so the two
measurements together can give tight limits on both $\Omega_0$
and $\Omega_\Lambda$ individually
\cite{Efsetal98,Garetal98,Teg98,TegEisHu98,Whi98,Lin98}.

As the bottom panels of Figure \ref{fig:models} show, variations to
$\Omega_0$ and $h$ affect the the height and width of the first
acoustic peak; the dependence is more precisely on
the quantity $\Omega_0 h^2$.  Thus, if the Hubble constant is
known, then the CMB can determine $\Omega_0$ and $\Omega$ (from
the peak location) and therefore the cosmological constant
$\Omega_\Lambda$.

A cosmological constant may also be distinguished from the CMB
via the additional large-angle anisotropy it produces via the
ISW effect \cite{KofSta86}  from density perturbations at
redshifts $z\la$few.  If there is a cosmological constant, there
should be a 
cross-correlation between the CMB temperature and some tracer of
the mass distribution, e.g.\ the extragalactic X-ray background
\cite{BouCriTur98} or weak lensing \cite{ZalSel99}, at these
redshifts \cite{CriTur96} (the same also occurs in an open
Universe \cite{Kam96,KinKam98}).  An experimental upper limit to 
the amplitude of this cross-correlation \cite{BouCriTur98} can
already be used to constrain $\Omega_0$, with some assumptions
about the bias of sources that give rise to the extragalactic
X-ray background.  If $\Omega_0\simeq0.3$ (either in an open or
a flat cosmological-constant model), then these X-ray sources
can be no more than weakly biased tracers of the mass
distribution \cite{KinKam98}.

\subsection{Rolling Scalar Fields}

The supernova evidence for an accelerating expansion has
engendered a burst of theoretical activity on exotic forms of
matter with an equation of state $p<-\rho/3$ (i.e.\ the equation of state needed
for $q_0<0$).  The simplest possibility is of course a
cosmological constant.  However, as explained in
Section \ref{sec:scalarfields}, a rolling scalar field
may also provide  
such an equation of state, provided the scalar field is not
rolling too quickly.  An almost endless variety of equations of
state---and expansion histories---are possible  in principle, 
given the freedom to choose the scalar-field potential and the
initial conditions.  This idea is variously referred to in the literature
as rolling-scalar-field, variable-cosmological-constant,
x-matter, generalized-dark-matter,
loitering-Universe, and/or quintessence models
\cite{RatPee88,SahFelSte92,SugSat92,Frietal95,CobDodFri97,
SilWag97,TurWhi97,ChiSugNak97,CalDavSte98,ChiSugNak98}.
Additional work has explored attractor solutions based on
exponential potentials 
\cite{LucMat85,Wet88,WanCopLid93,FerJoy97,CopLidWan98,LidSch99} or
``tracker-field'' solutions \cite{ZlaWanSte99,SteWanZla99} that
attempt to explain why the matter density would be comparable to
a scalar-field energy density today.

Because the expansion rate at decoupling in such models is the
same as in cosmological-constant models with the same
$\Omega_0$, the peak structure in the CMB is virtually
indistinguishable from that in cosmological-constant models
\cite{Hueetal99}.  However, perturbations in the scalar field
track perturbations to the matter density on large scales in
such a way that the large-angle ISW effect that appears in
cosmological-constant models is canceled by the effect of
scalar-field perturbations \cite{CalDavSte98}.
Data from cosmological observations, particularly
supernova measurements of the expansion history and
measurements of the power spectrum through large
galaxy surveys, may in principle
be used to break these degeneracies
\cite{Huetal99,Wanetal99}.

\section{Other Constraints on Particle Physics}

\begin{figure}[htbp]
\centerline{\psfig{file=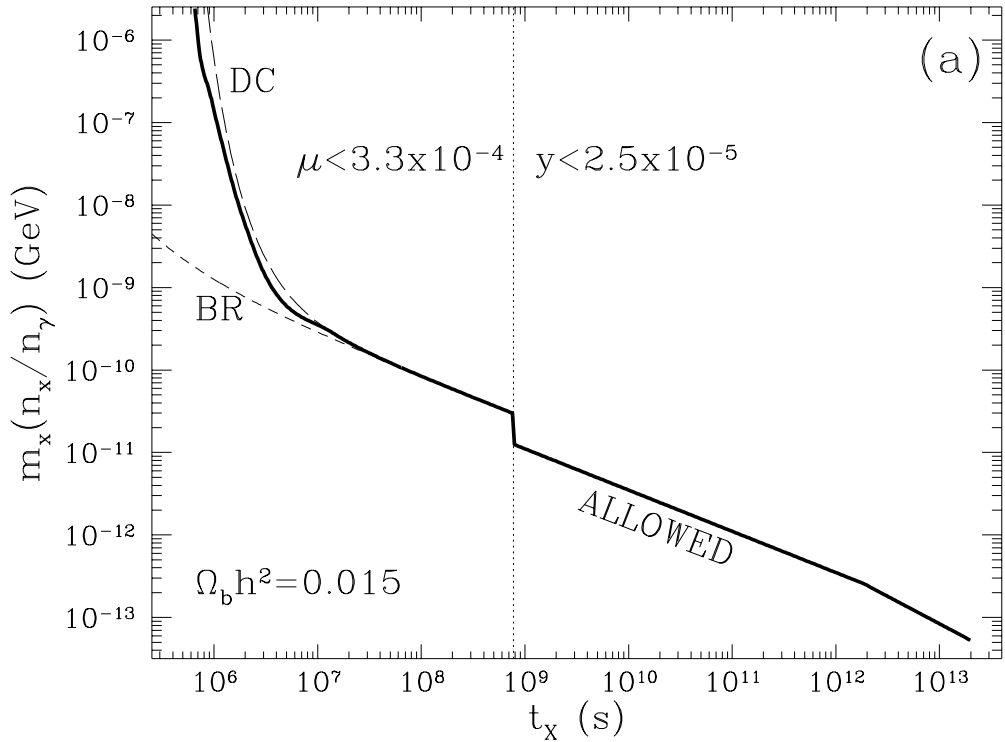,width=5.5in,angle={-90}}}
\bigskip
\caption{Constraints to the mass-lifetime plane for particles
         decaying to photons from FIRAS constraints to
	 distortions to the CMB blackbody spectrum.  \textit{Solid curve} is 
	 the numerical result; \textit{dashed curves} show various
	 approximations.  The quantity $n_X/n_\gamma$ is the
 	 initial ratio of the particle number density to the
 	 photon number density.  (From Reference
	 \cite{HuSil93a}.)}
\label{fig:decaylimits}
\end{figure}

\subsection{Decaying Particles}

As discussed in Section \ref{sec:freqspectrum},
FIRAS limits to  $\mu$ and $y$ distortions limit the injection
of energy into the early Universe and can thus be
used to constrain the mass-lifetime plane of particles that
decay to electromagnetically interacting particles (as shown in
Figure~\ref{fig:decaylimits}) \cite{HuSil93a,Elletal92}.
The CMB power spectrum can also constrain decaying particles.
For example, a neutrino of mass $\ga10$ eV that decays to
relativistic particles with a lifetime $\tau \simeq 10^{13-17}$
alters the expansion rate of the Universe between recombination
and today and thus produces large-angle anisotropy (via the ISW
effect) in disagreement with observations 
\cite{Lopetal98b,Han98b,Han98c}.

\subsection{Time Variation of Fundamental Constants}

A number of ideas for new physics postulate that some of the
fundamental constants of nature, such as the fine-structure
constant $\alpha$, may actually be varying (for a review, see
\cite{VarPot95}). Such a variation
could be caused by the cosmological evolution
of compact spatial dimensions in string theory
or Kaluza-Klein theories \cite{Mar84,Bar87,DamPol94}
or through scalar fields coupled to electromagnetism \cite{Car98}.
Limits of $|\Delta\alpha/\alpha| \la 10^{-7}$ were provided by
the natural nuclear reactor at Oklo \cite{Shy76,DamDys96},
and observations of atomic- and molecular-line positions at
high redshifts \cite{Sav56} provide limits of
$|\Delta\alpha/\alpha| < 3\times 10^{-6}$ at redshifts less than
1 \cite{Drietal98} and $|\Delta\alpha/\alpha| \la 3\times
10^{-4}$ at redshifts of 3 \cite{CowSon95,IvaPotVar98}. In fact, 
a detection of $\Delta\alpha/\alpha = -1.9\pm 0.5 \times
10^{-5}$ has been claimed on the basis of absorption lines at
redshifts greater than 1 \cite{Webetal99}, but there are some potential
problems with this result 
\cite{IvaPotVar98}.  Primordial nucleosynthesis can also provide
a less useful model-dependent limit \cite{KolPerWal86}. 

A change in $\alpha$ would affect the recombination
rate of hydrogen and thus alter the redshift of
last scatter.
This effect on the CMB can potentially lead to
upper limits on
$|\Delta\alpha/\alpha|$ between 0.01 and 0.001
\cite{KapSchTur98,Han98a} out to redshifts $z\simeq1100$, much
larger than those probed by quasar absorption spectra. 

\subsection{Topology of the Universe}

The fundamental cosmological assumptions of homogeneity and
isotropy require the Universe to be either the open, closed, or
flat Friedmann-Robertson-Walker model.  However, if the
assumption of isotropy is incorrect, then the Universe may
have some nontrivial topology (see~\cite{LacLum95}
for a review).
The open and flat FRW models have infinite volume, but a
Universe with either zero or negative curvature can have finite
volume if the Universe has nontrivial topology.  A number
of (somewhat imprecise) theoretical arguments suggest that a finite
Universe is easier to explain than an open Universe
\cite{ZelSta84} or could be used to explain the homogeneity of
the Universe \cite{Got80,EllSch86}.

If the volume of such a Universe is comparable to or less than that
observable today, then there may be signatures in the 
CMB.  Consider the simplest nontrivial topology (for a
flat Universe), that of a toroid.  If the Universe is a
three-dimensional toroid, then two different directions on the
sky will point to the same point in space, and there should be
observable correlations between the CMB temperature at distant
locations on the sky. Such models have essentially been
ruled out by \COBE\
\cite{FanMo87,Sok93,Sta93,Fan93,SteScoSil93,JinFan94,CosSmo95,CosSmoSta96,ScaLevSil98,LevScaSil98}. 

Interest in negative-curvature models with nontrivial topology
has reawakened recently because evidence seems to suggest
$\Omega_0\simeq0.3<1$, and thus possibly an open Universe.
If the Universe is negatively curved (hyperbolic), the spacetime 
volume element increases rapidly with distance, so that even if
the volume of the Universe is close to the horizon volume,
many copies of the Universe may still fit inside the
horizon volume.  Thus, none of the flat-Universe limits
on topology apply to hyperbolic
Universes \cite{CorSpeSta98b}. Furthermore, if the total density of
the Universe is $\Omega\simeq0.3$, the curvature scale is
small enough so that a huge number of topologies exist that
have proper volumes significantly smaller than the proper Hubble
volume \cite{CorSpeSta98c}.

Because the surface of last scatter is spherical, matched pairs of 
temperature circles would
appear in a negatively-curved Universe
with nontrivial topology provided that the topology radius were
smaller than the current horizon
\cite{CorSpeSta96,Wee98,CorSpeSta98c}.
Levin et al \cite{Levetal98}
propose searching for specific correlations between a given
pixel and all others in a map.  A null search for such
correlations in the \COBE\ maps ruled out a particular horn
topology \cite{Levetal97}.  Souradeep et al \cite{SouPogBon98} claim that
the \COBE\ maps already rule out most hyperbolic Universes
through this technique, although details have not been
presented.

\subsection{Primordial Magnetic Fields}

Magnetic fields of strength $10^{-6}$ G are
ubiquitous in our Galaxy and in distant clusters of galaxies.
All mechanisms for the origin of these magnetic fields postulate 
that they grew via some mechanism (e.g.\ dynamo or adiabatic
compression) from small
primordial magnetic fields.  However, the origin of these
primordial seed fields remains a mystery.  Many of the most
intriguing hypotheses the origin of these fields come from
new ideas in particle theory.  Proposed generation mechanisms
include inflation
\cite{TurWid88,CarFie91,GarFieCar92,Rat92,Dol93,GasGioVen95a,GasGioVen95b},
the electroweak \cite{Vac91,EnqOle93} or QCD phase transitions
\cite{QuaLoeSpe89,CheOli94}, a ferromagnetic Yang-Mills vacuum
state \cite{EnqOle94}, charge asymmetry \cite{DolSil93}, and
dilaton evolution \cite{Gio97}. 

Magnetic fields have several potentially
measurable effects: Faraday rotation \cite{KosLoe96} (A Mack,
A Kosowsky, manuscript in preparation) and
and associated depolarization \cite{HarHayZal96} of the
original CMB polarization; magnetosonic
waves that modify the acoustic oscillation frequencies
\cite{Adaetal96}; and Alfven waves, which can amplify vector
perturbations and induce additional correlations
\cite{DurKahYat98}, and for which diffusion 
damping is decreased, thereby increasing CMB power at
small scales \cite{SubBar98}.
The Faraday rotation signals can be detected through
the CC, TC, and GC 
power spectra they induce \cite{ScaFer97}
(although these power spectra are
frequency dependent).
A recent analysis of the \COBE\ maps has placed a limit on
a homogeneous primordial field strength corresponding
to $B_0< 3.4\times 10^{-9}(\Omega_0h_{50}^2)^{1/2}$ G
\cite{BarFerSil97} by searching for the temperature pattern of
a Bianchi type VII anisotropic spacetime
\cite{Nov68,BarJusSon85}.

\subsection{Large-Scale Parity Violation}

It is usually assumed that gravity is parity-invariant.
However, weak interactions are parity-violating
\cite{LeeYan56,Wu57}, and we surmise that the electroweak
interactions are united with gravity at the Planck scale by some
fundamental unified theory.  Are there any
remnants of parity-violating new physics in the early Universe?
As discussed in Section \ref{sec:powerspectra}, if either of
the temperature-polarization cross-correlation moments $C_l^{\rm
TC}$ or  $C_l^{\rm TG}$ is nonzero, it signals cosmological
parity breaking.
Lue et al and Lepora \cite{LueWanKam98,Lep98}
discuss how a parity-violating term, $\phi F_{\mu\nu}\tilde F^{\mu\nu}$
\cite{CarFie90,CarFie91,Car98}, that
couples a scalar field $\phi$ to the pseudoscalar ${\vec E}\cdot 
{\vec B}$ of
electromagnetism could yield nonzero $C_l^{\rm TC}$ and
$C_l^{\rm GC}$. Lue et al \cite{LueWanKam98} also discuss a
parity-violating term in the Lagrangian for gravitation that
would yield an asymmetry between the density of right- and
left-handed gravitational waves produced during inflation; such
an asymmetry would also give rise to nonzero $C_l^{\rm TC}$ and
$C_l^{\rm GC}$.  These parity-breaking effects would produce
frequency-independent $C_l^{\rm TC}$ and $C_l^{\rm GC}$,
unlike the frequency-dependent effect of Faraday rotation.

\subsection{Baryon Asymmetry}

There are very good theoretical and observational reasons to
believe that
our entire observable Universe is made of baryons and no
antibaryons.  But suppose momentarily that the observable
Universe consisted of some domains with
antibaryons rather than baryons (see e.g.\ \cite{Ste81}). 
If so, then particle-antiparticle 
annihilations at the interfaces of the matter and antimatter
regions would release a significant amount of energy
in $\gamma$-rays, thus heating the region and causing a
$y$-distortion of the CMB spectrum of order $y\simeq 10^{-6}$
\cite{KinKolTur97,CohRujGla98}. These distortions would appear in
thin strips on the sky which could potentially be
identified. However, the point is moot because limits on
the diffuse extragalactic $\gamma$-ray background limit the size of
our matter domain to be essentially as large as the horizon
\cite{CohRujGla98}.

\subsection{Alternative Gravity Models}

We now know through a variety of experiments
that general relativity provides an accurate accounting of
observed gravitational phenomena.  On the other hand, string theories
generically predict at least some small deviation from general
relativity, often in the form of scalar-tensor theories of
gravity \cite{Ber68,Nor70,Wag70,Bek77,BekMei78}.  The simplest
of these is 
Jordan-Fierz-Brans-Dicke (more commonly,
Brans-Dicke) theory
\cite{Jor49,Fie56,Jor59,BraDic61,Dic62,Dic68}.  An inflation
theory (``extended inflation'') based on Brans-Dicke gravity
\cite{LaSte89a,LaSte89b} was ruled out by the isotropy of the
CMB \cite{Wei89,LaSteBer89}, although models based on more
complicated scalar-tensor theories (``hyperextended inflation'') 
have also been considered
\cite{SteAcc90,BarMae90,GarQui90,HolKolWan90}.

Brans-Dicke theory includes a scalar field $\Phi$ and a new parameter 
$\omega$.  As $\omega \ra \infty$, the theory recovers general
relativity (in some sense).  Solar-system constraints from 
Viking spacecraft data limit $\omega \geq 500$ (for a
review, see \cite{Wil93}) and recent Very-Long Baseline Interferometry
measurements of time
delays of millisecond pulsars may further raise this limit
\cite{Wil98}.  In cosmological models based on Brans-Dicke
theories, general relativity is an attractor solution
\cite{DamNor93a,DamNor93b}, so gravity could have conceivably
differed from general relativity in the early Universe even if
it resembles general relativity today.  

Because the expansion rate and growth of gravitational-potential
perturbations are different in alternative-gravity theories, 
the precise predictions for CMB power spectra should be
different in these models.  
The epoch of matter-radiation equality is altered in
Brans-Dicke theories, and this may produce an observable signal
in forthcoming precise CMB maps \cite{LidMazBar98}.  
Cosmological perturbation
theory in scalar-tensor theories has been worked out
\cite{Nar69,PeeYu70,BapFabGon96,ChiSugYok98} and the CMB power
spectra calculated  
(X Chen, M Kamionkowski, manuscript in preparation).
If the scalar-field time
derivative $\dot\Phi$ is fixed to be small enough to be
consistent with big-bang-nucleosynthesis constraints
\cite{KamTur90,DamGun91,CasGarQui92,DamPic98} and
$\omega>500$, then the differences between the
general-relativistic and Brans-Dicke predictions is small,
although conceivably detectable with the Planck Surveyor.
Of course, the scalar-field evolution may be significantly
different in more sophisticated scalar-tensor theories, but
predictions for these models have yet to be carried out.

\subsection{Cosmic Rays}

We close this tour of the CMB/particle intersection
with possibly the oldest and most venerable
connection between these two topics.  Soon after the initial
discovery of the CMB, it was realized that cosmic rays
with energy $E\ga5\times 10^{19}$ eV can scatter from CMB
photons and produce pions. If a cosmic ray
is produced with an energy above $5\times 10^{19}$ eV, repeated
scatterings will reduce its energy to below this threshold
within a distance of about 50 Mpc
\cite{Gre66,ZatKuz66,Cro92,ElbSom95} (the
Greisen-Zatsepin-Kuzmin bound).
These constraint have become increasingly intriguing recently, as
several cosmic rays with energies $>10^{20}$
eV have been observed \cite{Lin63,Biretal94,Hayetal94,Biretal95}, and they do not appear to be coming from any
identifiable astrophysical sources (e.g.\ radio 
galaxies or quasars \cite{Hil84}) as near as 50 Mpc
\cite{Hayetal94,Biretal95,ElbSom95,Bie97}.  So where are these
cosmic rays coming from?  Some possibilities are exotic
production mechanisms such as topological defects
\cite{Hiletal86,AhaBhaSch92,BhaHilSch92,Chietal93,
SigSchBha94} or supermassive unstable particles
(\cite{KuzRub98,BerKac98,BirSar98}; see \cite{SigBha98} for
a review).  If a recently claimed alignment of the
highest-energy events with very distant radio quasars
\cite{FarBie98} is confirmed by larger numbers of events, then it 
may be that these cosmic rays are exotic particles that interact 
with baryons but not photons \cite{ChuFarKol98,AlbFarKol99},
e.g.\ supersymmetric $S_0$ baryons \cite{Far84,Far95,Far96}.
In the absence of any compelling traditional astrophysical
origin, it seems that the simultaneous existence of the CMB and
these cosmic rays may be pointing to some intriguing new
particle physics.

\section{Summary, Current Results, and Future Prospects}

The primary cosmological observables pursued by CMB experiments
are the frequency spectrum of the CMB, parameterized by
$\mu$ and $y$ distortions, and the angular temperature and
polarization power spectra, $C_l^{\rm TT}$, $C_l^{\rm GG}$,
$C_l^{\rm CC}$, $C_l^{\rm TG}$, $C_l^{\rm TC}$, and $C_l^{\rm
GC}$.  There are additional observables, such as higher-order
correlation functions or cross-correlation of the CMB
temperature/polarization with other diffuse extragalactic
backgrounds.  Rough
estimates of the $C_l^{\rm TT}$ at degree angular scales were
obtained \cite{WhiScoSil94} from the first generation of ground-based and
balloon-borne CMB experiments. Forthcoming
experiments will require far more sophisticated techniques for
disentangling the CMB from foregrounds, and for recovering the power
spectra from noisy data and from maps that cover only a fraction
of the sky.  A large literature 
is now devoted to these important issues,
which we cannot review here.

Progress in CMB experiments is so rapid at the time
of writing that any current data we might review would
almost certainly become obsolete by the time of publication.
We therefore refrain from showing any experimental results
in detail and instead describe the current observations qualitatively.
First, there is the isotropy of the CMB, which has long been
explained only by
inflation.  Among the numerous pre-\COBE\ models for the origin
of large-scale structure, those based on a nearly
scale-free spectrum of primordial adiabatic perturbations seem
to account most easily for the amplitudes of both the
large-angle CMB anisotropy measured by \COBE\ and the amplitude
of clustering in galaxy surveys.  The galaxy distribution seems
to be consistent with primordial Gaussianity.

Moreover, data from a large number of CMB experiments that probe
the angular power spectrum at degree angular scales have now
found (fairly convincingly) that there is significantly more
power at degree angular separations ($l\sim200$) than at \COBE\
scales, as one would expect if the acoustic peaks do
exist, but in apparent conflict with most theorists'
expectations for the degree-scale anisotropy in
topological-defect models.
The
existence of this small-scale anisotropy further suggests no
more than a small level of reionization (i.e.\ $\tau\ll 1$).  At
the time of writing, the measurements are not precise
enough to discern either the first or any higher peaks in the
temperature power spectrum (some recent data are shown in
Reference \cite{Kam98} and are usually updated at Reference \cite{Teg99}).
Some experiments have claimed to see the outline of a first
acoustic peak at $l\sim200$ \cite{Petetal99} (which would
indicate a flat Universe).  Moreover, some maximum-likelihood
analyses of combined results of all experiments claim that the
data indicate a flat Universe \cite{LinBar98,Hanetal98}.
However, these results are not yet robust.

Thus, although inflation is by no means yet in the clear,
observations do seem to be pointing increasingly toward
inflation.  MAP and the Planck Surveyor will soon make far more
precise tests of inflation (see \cite{Kam98} for simulated
data from MAP and Planck).  First of all, the predictions of
primordial adiabatic perturbations will be tested with
unprecedented precision by the  peak structure in the CMB
temperature power spectrum.  If the peaks do appear, then MAP
and the Planck Surveyor should be able to measure the total
density $\Omega$ to a few percent or better \cite{Junetal96a} by
determining the location of the first acoustic peak
\cite{KamSpeSug94}.  Moreover, by fitting MAP and Planck
satellite data to theoretical curves, such as those shown in
Figure \ref{fig:models}, precise information on the values of
other classical cosmological parameters can also be obtained
\cite{Junetal96b,BonEfsTeg97,ZalSelSpe97,DodKinKol97,BonEfs98}.
If nonrelativistic matter outweighs baryons, then it should
become evident with MAP and Planck.  The existence of a
cosmological constant will further be tested, and some of the
tests of gravity, decaying particles, etc, that we
have reviewed will become possible.

If MAP and Planck confirm that the Universe is flat and that
structure grew from primordial adiabatic perturbations, then the next 
step will be to search for the gravitational-wave background
predicted by inflation.  Such a gravitational-wave background
could be isolated uniquely with the curl component of the
polarization.  If the inflaton-potential height is $V^{1/4}\ll
10^{15}$ GeV, then the gravitational-wave background will be
unobservably small.  However, if inflation had something to do
with grand unification (i.e.\ $V^{1/4} \sim 10^{15-16}$~GeV, as
many theorists surmise), then the curl component of the
polarization is conceivably detectable with the Planck Surveyor
or with a realistic next-generation dedicated polarization
satellite experiment.  If detected, the curl component would
provide a ``smoking-gun'' signature of inflation and indicate
unambiguously that inflation occurred at $T\sim10^{15-16}$ GeV.
Although an observable signature is by no means guaranteed, even 
if inflation did occur, the prospects for peering directly back to
$10^{-40}$ sec after the big bang are so
tantalizing that a  vigorous pursuit is certainly warranted.

\bigskip

\leftline{\textsc{Acknowledgments}}
We thank R Caldwell, A Liddle, and L Wang for very 
useful comments.
MK was supported by a DOE Outstanding Junior Investigator
Award, DE-FG02-92ER40699, NASA Astrophysics Theory Program grant
NAG5-3091, and the Alfred P. Sloan Foundation.  AK was supported
by NASA Astrophysics Theory Program grant NAG5-7015 and
acknowledges the kind hospitality of the Institute for Advanced
Study.

{\twocolumn

}
\end{document}